\documentclass[prb,preprint,aps,superscriptaddress,longbibliography]{revtex4-1}

\usepackage{graphicx, epstopdf}
\usepackage{subfigure}
\usepackage{float}
\usepackage{amsmath}
\usepackage{amsfonts}
\usepackage{amssymb}
\usepackage{bm}
\usepackage{subfigure}
\usepackage{caption}
\usepackage{hyperref}

\usepackage{color,soul}
\usepackage{xcolor}
\usepackage{hyperref}
\hypersetup{
     colorlinks = true,
     citecolor  = red,  
     linkcolor  = blue  
}

\usepackage{tabularx} 
\usepackage{multirow}

\begin{document}

\title{Topological Semimetals Predicted from First-principles Calculations}

\author{Hongming Weng}
\email{hmweng@iphy.ac.cn}
\affiliation{Beijing National Laboratory for Condensed Matter Physics,
      and Institute of Physics, Chinese Academy of Sciences, Beijing
      100190, China}
      
\affiliation{Collaborative Innovation Center of Quantum Matter,
  Beijing, China}

\author{Xi Dai}
\email{daix@iphy.ac.cn}
\affiliation{Beijing National Laboratory for Condensed Matter Physics,
      and Institute of Physics, Chinese Academy of Sciences, Beijing
      100190, China}
      
\affiliation{Collaborative Innovation Center of Quantum Matter,
  Beijing, China}

\author{Zhong  Fang}
\email{zfang@iphy.ac.cn}
\affiliation{Beijing National Laboratory for Condensed Matter Physics,
      and Institute of Physics, Chinese Academy of Sciences, Beijing
      100190, China}
      
\affiliation{Collaborative Innovation Center of Quantum Matter,
  Beijing, China}


\begin{abstract}
  We have given a summary on our theoretical predictions of three
  kinds of topological semimetals (TSMs), namely, Dirac semimetal
  (DSM), Weyl semimetal (WSM) and Node-Line Semimetal (NLSM).  TSMs
  are new states of quantum matters, which are different with
  topological insulators.  They are characterized by the topological
  stability of Fermi surface, whether it encloses band crossing point,
  i.e., Dirac cone like energy node, or not.  They are distinguished
  from each other by the degeneracy and momentum space distribution of
  the nodal points. To realize these intriguing topological quantum
  states is quite challenging and crucial to both fundamental science
  and future application. In 2012 and 2013, Na$_3$Bi and Cd$_3$As$_2$
  were theoretically predicted to be DSM, respectively. Their
  experimental verifications in 2014 have ignited the hot and
  intensive studies on TSMs. The following theoretical prediction of
  nonmagnetic WSM in TaAs family stimulated a second wave and many
  experimental works have come out in this year.  In 2014, a kind of
  three dimensional crystal of carbon has been proposed to be NLSM due
  to negligible spin-orbit coupling and coexistence of time-reversal
  and inversion symmetry. Though the final experimental confirmation
  of NLSM is still missing, there have been several theoretical
  proposals, including Cu$_3$PdN from us. In the final part, we have
  summarized the whole family of TSMs and their relationship.
\end{abstract}

\maketitle

\noindent
{\bfseries Table of Contents}\medskip


\noindent
{1.} Introduction\\

\noindent
{2.}   Dirac Semimetal\\
\hspace*{12pt}{2.1:}  Na$_3$Bi \\
\hspace*{12pt}{2.2:}  Cd$_3$As$_2$ \\

\noindent
{3.}   Weyl Semimetal\\
\hspace*{12pt}{3.1:}  Magnetic Weyl Semimetal: HgCr$_2$Se$_4$ \\
\hspace*{12pt}{3.2:}  Non-magnetic Weyl Semimetal: TaAs family\\

\noindent
{4.}   Node-Line Semimetal \\
\hspace*{12pt}{4.1:}   All Carbon Mackay-Terrones Crystal \\
\hspace*{12pt}{4.2:}   Anti-Perovskite Cu$_3$PdN \\

\noindent
{5.}    Discussion and Prospect\\

\noindent
{6.}    Acknowledgement \\

\noindent
{7.}    References \\


{\large\bf 1. Introduction}
\hspace*{2pt}

Classification of electronic states is one of the key concepts in
condensed matter physics. An usual way to classify the electronic
states in solids is based on the symmetry principle, by which most of
the states as well as the transitions among them can be
understood. For example, the ferro-electric and ferro-magnetic states
can be understood as states that break the spacial inversion and time
reversal symmetries, respectively. The Landau theory for phase
transitions is very successful in condensed matter physics, which is
built entirely on the symmetry principle. In the recent thirty years,
another way to classify condensed matter systems has been developed
through a completely different point of view, i.e. topology. The first
type of systems which can be classified by its topological features is
the 2D electron gas under high magnetic field. Once there are integer
number of Landau levels being fully occupied, the system is in $n$-th
integer quantum Hall state (IQHS) with quantized Hall
conductance.~\cite{qhe} Further analysis of IQHS by TKNN revealed that
the IQHS has very deep topological origin.~\cite{tknn, laughlin}
Although different IQHS are the same in terms of symmetry, they can be
characterized by different integers (TKNN number or Chern number),
which can be expressed by the integral of the Berry's curvature over
the whole magnetic Brillouin zone (BZ).~\cite{tknn} This integer
number is called topological invariance, which can not be changed
without closing the energy gap of the system.

After the discovery of IQHS, it was anticipated that the concept of
topological electronic states should not be limited only to 2DEG under
external magnetic field~\cite{haldane_model_1988}. The idea ought to
be generalized to the realistic materials as well. In about ten years
ago, the idea of finding electronic states that can be classified by
topological invariance in solid state materials has been realized in a
class of band insulators, topological insulator.~\cite{Kane_Z2_2005,
  bernevig_quantum_2006,Hasan_Kane_RMP_2010,QiXL_RMP_2011} Similar
with the IQHS, the TI can be characterized by topological invariances
or indices but the difference is that these indices, called Z2
indices, can only take two possible value, even or odd. The TI,
characterized by odd Z2 indices, has unique Dirac like surface (or
edge) states, which is unavoidable as long as the time reversal
symmetry is preserved. The first TI proposed by theorists is
graphene~\cite{Kane_Mele_Graphene_PRL_2005}, while due to its
extremely small energy gap (about 10$^{-3}$ meV) opened by spin-orbit
coupling,~\cite{PhysRevB.75.041401} it hard to observe any physical
properties of TI in graphene. The prediction of TI state in HgTe/CdTe
quantum well was made by the Stanford
group~\cite{Bernevig_HgTe_QW_science_2006} and had been confirmed
experimentally a year after.~\cite{Molenkamp_HgTe_science_2007} The
discovery of 3D TI Bi$_2$Se$_3$ family~\cite{ZhangHJ_Bi2Se3_2009NP,
  TI_exp_Hasan_2009, chen_experimental_2009} makes it much easier to
study all kinds of properties of TI including the transport under
magnetic field, thermal electric
properties~\cite{XuYong_PRL2014,thermal_Bi2Se3} as well as the
possible topological superconductor phase induced by proximate
effect.~\cite{LiangFu_MF_TI_2008PRL, QiXL_MF_from_QHE_2010PRB,
  zhang_pressure-induced_2011} With Doping of the magnetic elements
into the Bi$_2$Te$_3$ family thin films, another long awaited
topological phase, quantum anomalous Hall state
(QAHE)~\cite{haldane_model_1988, onoda_quantized_2003, HgMnTe,
  qahe_advphy}, has been theoretically
predicted~\cite{yu_quantized_2010} and experimentally
observed.~\cite{chang_experimental_2013} The QAHE can be viewed as a
kind of IQHS without Landau Levels (no external magnetic
field).~\cite{qahe_advphy}

With the successful discovery of those topological insulating states,
it is nature to ask wether or not metals can be classified by
topological invariances as well.  Although till now the general
topological classification for metals is still unclear, we now have
ways to classify a special type of metals, the semimetals, where the
zero energy contour in energy dispersion or Fermi surface contains
only isolated points (or lines for 3D systems) rather than
surfaces. These Fermi nodes (or nodal lines) are caused by the band
crossing points (also called as nodes or nodal points) right at the
Fermi level. The semimetal can be viewed as the special type of
``insulator'' as well, where the energy gap only closes at those
isolated $k$-points or lines. In 2D systems, graphene can be viewed as
the first well studied topological semimetal (TSM), where the low
energy physics can be well described by the 2D massless Dirac
equation.  The line integral of the Berry connection along any closed
loops in the BZ will accumulate a phase $\pi$ ($0$) if the loop
(doesn't) encloses the Dirac point and this phase can be used as the
topological invariance, which manifests itself in the unique Landau
level structure under magnetic field. While as we will introduce
below, similar to graphene the band crossing points can also occur in
3D materials.  Although the band crossings in 3D and 2D do share some
similarities, it is indeed very different. The most essential
difference comes from the factor that without extra degeneracy,
i.e. the spin degeneracy for fermions, the band crossings in 3D can
only be shifted but not removed by small perturbations in
Hamiltonian. While in comparison, the band crossing points in 2D is
caused by crystalline symmetry on some special $k$-points in the BZ
and will disappear immediately once a small perturbation is added to
break that symmetry.  Therefore in 3D, the crossing points between two
non-degenerate bands can be viewed as ``topological defects", which
are the stable objects like the vortices in the fluid. Such crossing
points are called Weyl points (Weyl nodes) because the low energy
physics around them can be well described by the Weyl equation, which
is well known in particle physics and contains only half of the
freedom of the Dirac equation.~\cite{weyl_elektron_1929} Weyl nodes in
crystalline solids were first studied as magnetic monopoles in
momentum space in the context of anomalous Hall
effect~\cite{fang_anomalous_2003} As will be illustrated in detail
below, Weyl points have chirality defined as the sign of the
determinant for the velocity tensor, which is another unique point in
3D.  It has been proved mathematically that Weyl points generated in
any lattice model can only come in pairs with opposite
chirality.~\cite{nielsen_absence_1981-1,nielsen_absence_1981-2,
  Balents_Physics_Weyl_2011} In fact, the only way to remove Weyl
points is to move a pair of Weyl points with opposite chirality to a
same k point and annihilate them. The most important effect generated
by these ``chiral electrons" is so called ``chiral anomaly" describing
the adiabatic electron pumping caused by the joint effect of the
external magnetic and electric fields between Weyl points with
opposite chirality.~\cite{nielsen_adler-bell-jackiw_1983} The chiral
anomaly (or ABJ anomaly) will generate current along the direction of
the magnetic field leading to negative magneto resistance when the
magnetic and electric fields are parallel to each
other.~\cite{Son2013,Kim2014} Besides the negative MR, chiral anomaly
can also generate more effects in
transport~\cite{PhysRevLett.108.046602} and optical
properties,~\cite{PhysRevB.89.245121,Norman2015,WeylCD_Yu2015,
  WeylCd_Ran2015} as well as the chiral magnetic
effect~\cite{Transport_Weyl_XLQi_2013, Li2014}, nonlocal
transport,~\cite{nonlocalWeyl2014} and chiral gauge
anomaly.~\cite{chiralgauge2013}

\begin{figure}[tbp]
\begin{centering}
\includegraphics[clip,width=0.95\textwidth]{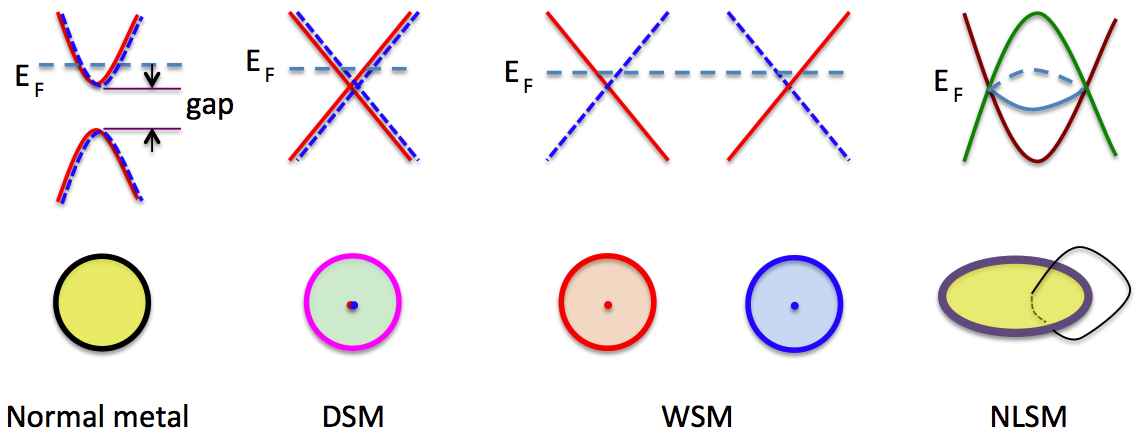} \par
\end{centering}
\caption{(Color online) Schematic band structure and Fermi surface (with Fermi level shift slightly off 
the nodal points) for normal metal and three kinds of topological semimetals, Dirac semimetal (DSM), Weyl semimetal (WSM) 
and Node-Line Semimetal (NLSM). The close path (black thin circle) interlocked with the nodal ring (thick circle) is also shown.
}
\label{fig:TSM_fermisurf}
\end{figure}

The nodal point behaves like the magnetic monopole in crystal momentum
space, being a ``source" or ``sink" of Berry curvature, the field of
Berry flux.~\cite{berry, fang_anomalous_2003, qahe_advphy}
Fig.~\ref{fig:TSM_fermisurf} shows the typical band structure and the
topological relationship between Fermi surface (with Fermi level
slightly off the node) and magnetic monopoles. The Dirac semimetal
(DSM) has its Fermi surface encloses two monopoles with opposite
topological charge ``kissing" at the same $k$-point, while WSM always
has opposite monopoles in pair and they are well
separated.~\cite{Balents_Physics_Weyl_2011} Node-Line semimetal (NLSM)
is a special case with nodal points forming a periodically continuous
line or closed ring in the momentum space. After a loop of adiabatic
evolution along a closed path interlocked with the nodal ring,
eigenstates on one of the crossing bands will acquire a Berry phase of
$\pi$, which stabilize its ring like Fermi surface. Therefore, these
three kinds of TSM can be identified by the degeneracy and
distribution of nodal points.

These exotic nontrivial TSMs are expected to have intriguing quantum
phenomena and physical properties.  Their realistic material
realization become very important and crucial to bring them into the
realm of experimental study and potential application in future. As we
have pointed out in Ref.~\onlinecite{MRS_review}, the quick
development of first-principles calculation methods and software
packages, together with the intrinsic robustness of band topology
immune to perturbation and numerical error, ensure the predictive
power of the stat-of-art first-principles calculation for the
topological materials. The remarkable successful stories include the
2D TIs~\cite{Bernevig_HgTe_QW_science_2006,
  Molenkamp_HgTe_science_2007}, 3D TIs~\cite{ZhangHJ_Bi2Se3_2009NP,
  chen_experimental_2009, TI_exp_Hasan_2009}, Chern insulators
supporting quantum anomalous Hall effect~\cite{yu_quantized_2010,
  chang_experimental_2013}, 3D topological crystalline
insulators~\cite{FuLiang_Topological_Crystalline_2011PRL,Hsieh_EXP_TCI_2012}
and so on.  In this topical review, we will briefly summarize our
theoretical material proposals for three types of TSMs, namely DSM,
WSM and NLSM. Some of them have been confirmed by experimental
observations after our predictions.  Experimentally, the most direct
and efficient way to observe the TSMs is to use the angle resolved
photoemission spectroscopy (ARPES) technique, which can detect the
band structure and even its spin polarization. There are basically
three important hallmarks of TSMs to be observed by ARPES, including
the 3D Dirac/Weyl points, the Fermi arcs on the
surface~\cite{WanXG_WeylTI_2011} and the spin texture of Fermi arcs.

{\large\bf 2. Dirac Semimetal}
\hspace*{2pt}


The Dirac equation describing electrons with relativistic effect~\cite{Dirac610} is written as  
\begin{eqnarray*}
\left(\begin{array}{cc}
      \hat{E}(\bf{k})-\mathbf{\sigma}\cdot\hat{\bf{k}} &  0 \\
      0 & \hat{E}(\bf{k})+\mathbf{\sigma}\cdot\hat{\bf{k}} 
\end{array}
\right)\psi=mc^2\left(\begin{array}{cc}
0 & I_2 \\
I_2 & 0 
\end{array}
\right)\psi.
\end{eqnarray*}
Here $m$ represents the mass term and couples the two massless Weyl fermions 
described by $\pm\sigma\cdot\bf{k}$ with opposite chirality.~\cite{weyl_elektron_1929,RevModPhys.47.331,volovik_universe_2009} 
If $m$ is finite, this equation describes the massive Dirac fermion as shown in Fig.~\ref{fig:TSM_fermisurf} 
assuming the Fermi level inside of the gap. If $m$ is zero, it describes the massless 
Dirac fermion composed by two massless Weyl nodes overlapping each other. If Fermi level 
is slightly shifted off the Dirac node, the Fermi surface encloses a pair of Weyl
nodes. The number of pairs can be looked as a topological invariance to identify DSM. 

\begin{figure}[tbp]
\includegraphics[clip,scale=0.6]{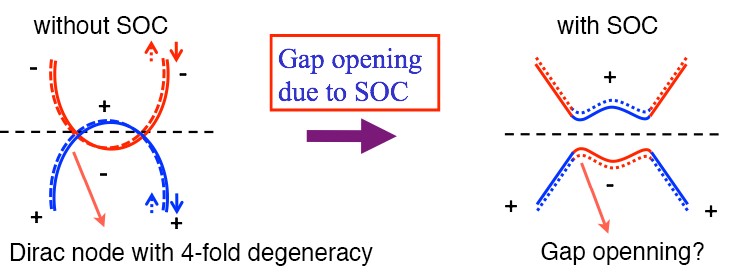}
\caption{(Color online) Schematic plot of band inversion mechanism. (left) The case 
without SOC, band inversion between two spin degenerate (dashed line for up spin, solid one 
for down spin) bands. + and - indicates the parity of the states at time-reversal invariant 
moment. (right) After including SOC, the band crossings are lifted and band gap is open. }
\label{fig:bandInv_DSM}
\end{figure}

In a general band inversion mechanism as shown in Fig.~\ref{fig:bandInv_DSM},
the band inversion between two spin degenerate bands can cause
four-fold degenerate energy nodal points. However, in general situation 
as discussed by S. Murakami,~\cite{murakami_phase_2007, Murakami2011748}
including SOC will open a band gap at the four-fold degenerate
band crossing points. This leads to insulating state and most probably
the system becomes a TI, or a topological crystalline insulator, or any other insulator.
However, we think such band gap opening is not inevitable
and it can be well protected by some crystal symmetry as long as the two spin degenerate
bands belong to two different 2D irreducible representations of the little group
at the $k$-point where the band crossing happens. Therefore, the candidate of DSM 
must have at least two different 2D irreducible representations in its double space 
group. This is a necessary condition but not enough one. This finding is very crucial
and it directly guides the theoretical prediction of DSM in Na$_3$Bi~\cite{WangZJ_Na3Bi_2012} 
and Cd$_3$As$_2$.~\cite{WangZJ_Cd3As2_2013}  

{\bf 2.1: Na$_3$Bi}
\hspace*{2pt}



\begin{figure}[tbp]
\begin{centering}
\includegraphics[clip,width=0.95\textwidth]{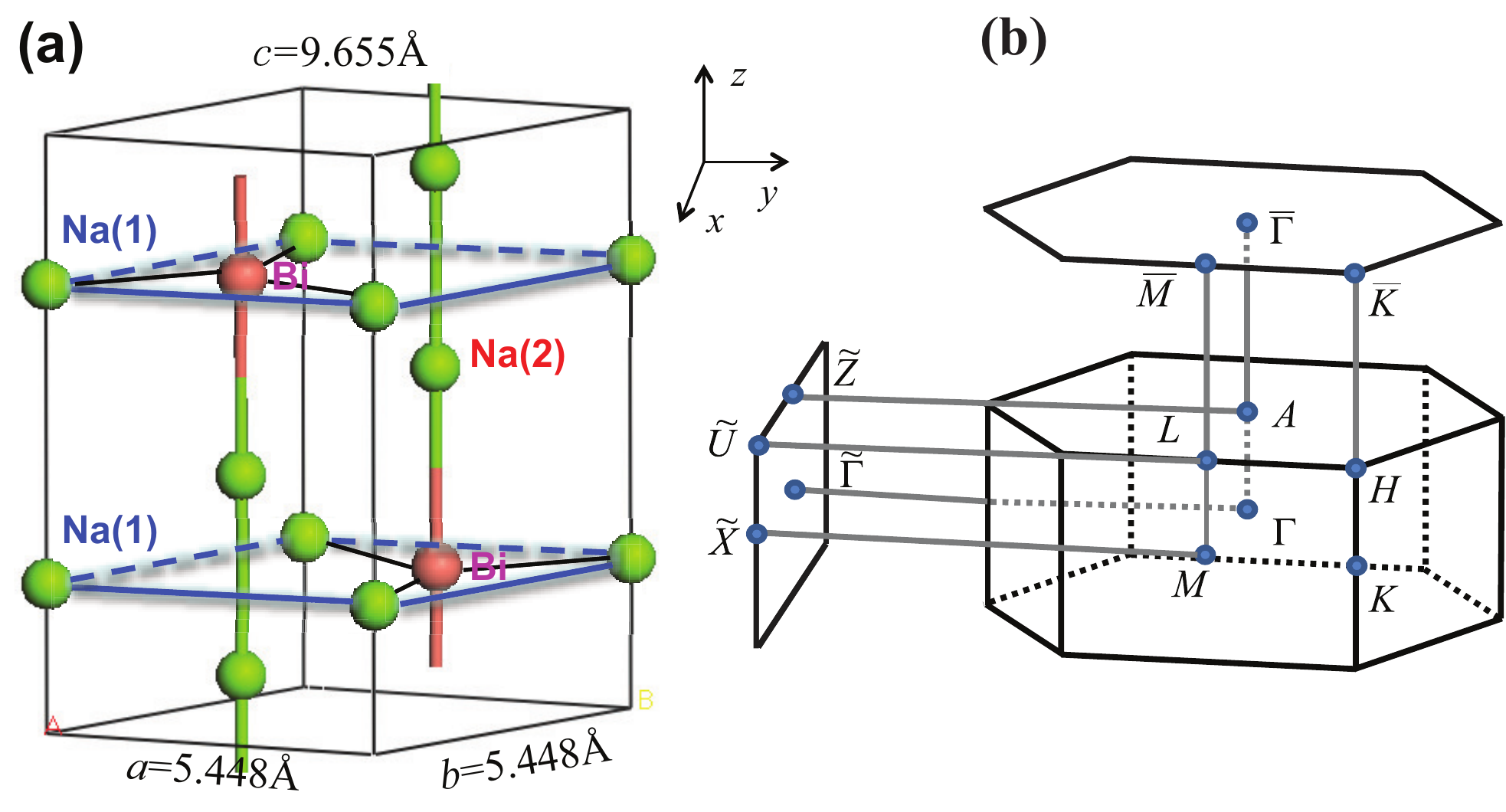} \par
\end{centering}
\caption{(Color online) (a) Crystal structure of Na$_3$Bi with
  $P6_3/mmc$ symmetry. Na(1) is at $2b$ position
  $\pm$(0,0,$\frac{1}{4}$), and Bi is at $2c$ position
  $\pm$($\frac{1}{3}$,$\frac{2}{3}$,$\frac{1}{4}$). They form
  honeycomb lattice layers. Na(2) are at $4f$ position
  $\pm$($\frac{1}{3}$, $\frac{2}{3}$,$u$) and
  $\pm$($\frac{2}{3}$,$\frac{1}{3}$,$\frac{1}{2}+u$) with $u$=0.583,
  threading Bi along $c$ axis. (b) Brillouin Zone of bulk and the
  projected surface Brillouin Zones of (001) and (010) plane.}
\label{fig:A3Bi-elec}
\end{figure}

In 2012, present authors and their collaborators predicted that DSM 
might be realized in some hexagonal phase of 
alkali pnictides $A_3B$ ($A$=Alkali metal, $B$=As, Sb or Bi) represented
by Na$_3$Bi.~\cite{WangZJ_Na3Bi_2012} Its crystal structure is shown in Fig.~\ref{fig:A3Bi-elec}.
There are two nonequivalent Na sites, Na(1) and Na(2), in one unit cell. 
Na(1) and Bi form simple honeycomb lattice layers stacking along 
the $c$-axis. Na(2) atoms are inserted at the interlayer position, connecting 
these layers on top of Bi atoms. From the ionic picture, due to the closed-shell configuration
where the number of valence electrons (3$\times$Na-$s^1$+Bi-$p^3$) is
equal to 6, one may expect Na$_3$Bi is a semiconductor, similar to 
Na$_3$Sb~\cite{Na3Sb}. However, the electronegativity 
of Bi is weaker than Sb, which makes Na$_3$Bi be DSM rather semiconductor 
like Na$_3$Sb.

\begin{figure}[tbp]
\includegraphics[clip,scale=0.6]{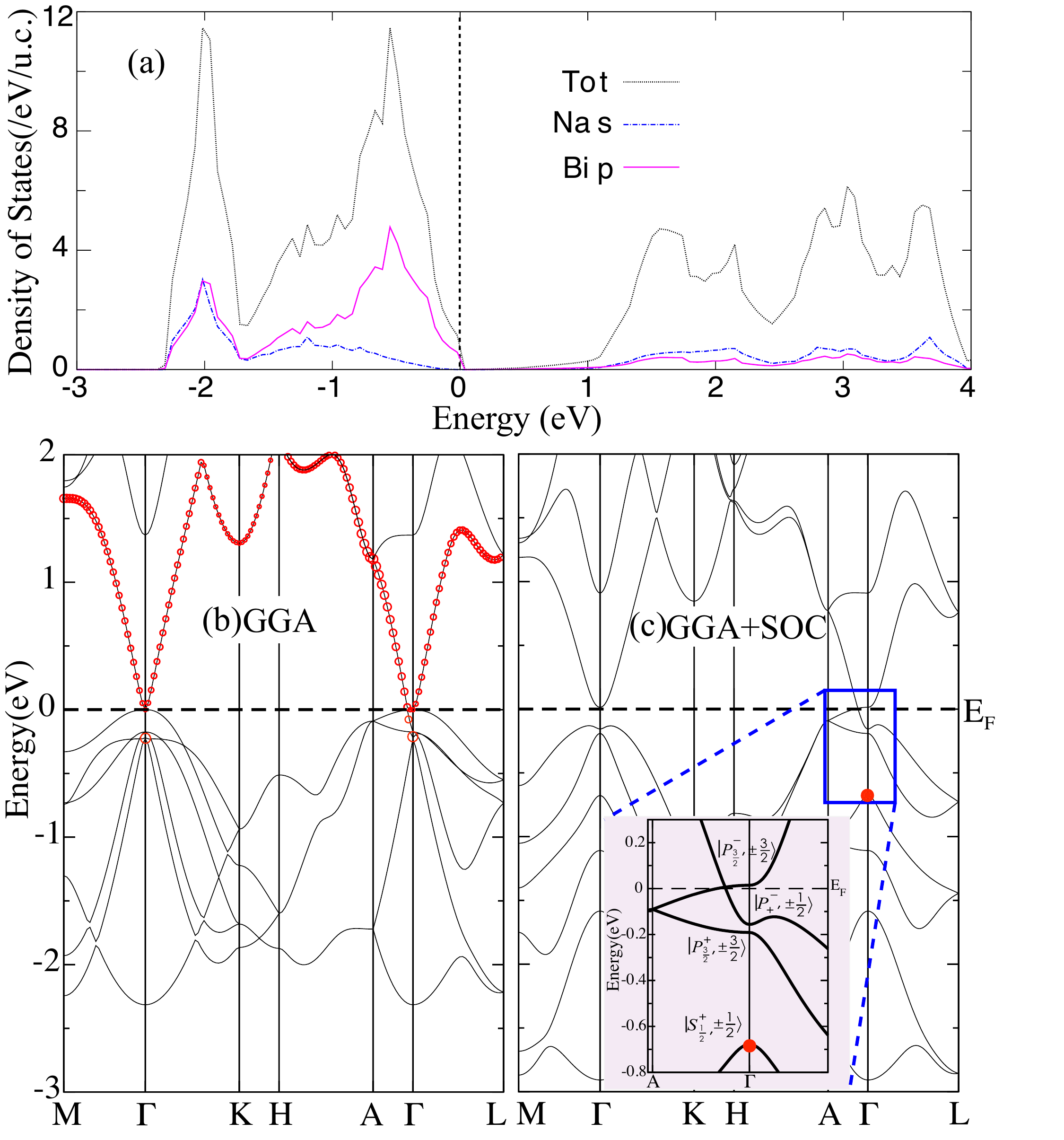}
\caption{(Color online) The calculated electronic structures of
  Na$_3$Bi. (a) The total and partial density of states. (b) and (c)
  are the band structures without and with spin-orbit
  coupling, respectively. The red circles indicate the projection to the Na-3$s$
  states. The orbital characters of wave-functions at $\Gamma$ point
  are labeled in the inset (see Effective Hamiltonian for details).}
  \label{fig:Na3Bi_band}
\end{figure}

The calculated band structures of Na$_3$Bi are shown in Fig.~\ref{fig:Na3Bi_band}.
They suggest that the valence and conduction bands are dominated by Bi-$6p$ 
and Na-$3s$ states. Around the Fermi level, the top valence band is mostly
from Bi-$6p_{x,y}$ states and the lowest conduction band is mostly 
from Na(1)-$3s$ orbital. These features are similar to those of 
Na$_3$Sb~\cite{Na3Sb} except that in Na$_3$Bi, the Na-$3s$ band is lower 
than Bi-$6p_{x,y}$ by about 0.3 eV at $\Gamma$, and this value is 
enhanced to be 0.7 eV in the presence of SOC. This is a typical picture of
band inversion~\cite{MRS_review} but absent in Na$_3$Sb. The band inversion is due to 
the heavier Bi, which has higher $6p$-states and larger SOC compared to Sb. 
Since GGA usually underestimates the band gap and results in the overestimation
of band inversion, the band inversion is further confirmed by the calculation using 
hybrid functional HSE~\cite{HSE03,HSE06}. The band inversion persisted and 
Na-$3s$ band is around 0.5 eV lower than Bi-$6p$ ones. K$_3$Bi and Rb$_3$Bi 
have the similar band structure and band topology with band inversion around 
0.33 eV and 0.42 eV, respectively. 

Different from topological insulators such as Bi$_2$Te$_3$ and
Bi$_2$Se$_3$,~\cite{ZhangHJ_Bi2Se3_2009NP, ZhangWei_NJP_BiSe_2010} 
Na$_3$Bi is a semimetal with two nodes (band-crossings) exactly at 
Fermi level as shown in Fig.~\ref{fig:Na3Bi_band}. 
Its Fermi surface consists of two isolated Fermi points, which are located
at (0, 0, $k_z^c$$\approx$$\pm$0.26$\times \frac{\pi}{c}$) along the
$\Gamma$-A line. Around each node, the band dispersions is linear, resulting in a 3D
Dirac cone. It is different from the Dirac cone in graphene not only in
dimensionality, but also in its robustness, because the Fermi points
here survive in the presence of SOC, but a tiny band gap open in graphene.~\cite{PhysRevB.75.041401}

\begin{figure}[tbp]
\includegraphics[clip,scale=0.6]{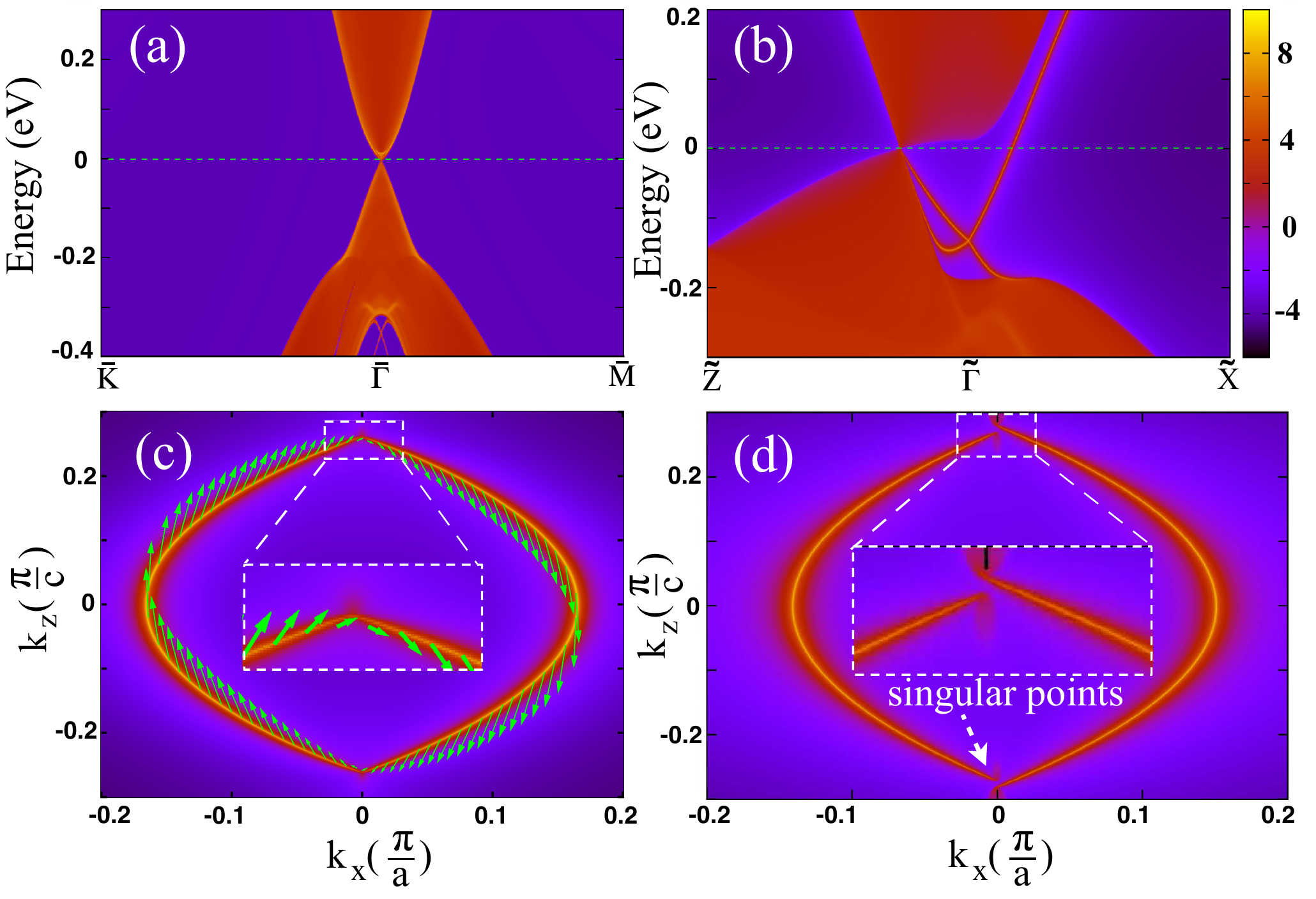}
\caption{(Color online) The projected surface states and their fermi
  surfaces of Na$_3$Bi. (a) and (b) are the projected surface density
  of states for [001] and [010] surfaces, respectively. (c) The fermi
  surfaces (fermi arcs) and its spin texture (in-plane component) for
  the [010] surface states. (d) The fermi arcs of [010] surface
  obtained from fitted effective hamiltonian with additional exchange
  field $h_1$=6 meV (see Phase diagram for details). The discontinuity
  around the singular fermi points becomes now obvious (enlarged in
  the insets).}
  \label{fig:Na3Bi_surfstate}
\end{figure}

The crossing bands along the $\Gamma$-A line belong to two different
irreducible representations distinguished by the three-fold rotational
symmetry around $\Gamma$-A axis. Breaking this symmetry will introduce interaction between
them and make it insulating. It has been tested that 1\% compression
along $y$ axis will open up a gap $\approx$5.6 meV. This insulating
state, however, is topologically non-trivial with
$Z_2$=1~\cite{Hasan_Kane_RMP_2010, QiXL_RMP_2011} due to the inverted band structure
around the $\Gamma$ point. This fact leads to coexistence of 
both bulk 3D Dirac cone and topological surface states (a single
pair) as long as the crystal symmetry protecting band crossings stands.
As shown in Fig.~\ref{fig:Na3Bi_surfstate}(a), the solid Dirac cone is
from the projection of bulk 3D Dirac cone onto [001]-surface overlapped
with the topological surface states. Such distinguished property of DSM is 
more obvious if inspecting its side surface on which two Dirac cones are 
projected at two different points. As shown in Fig.~\ref{fig:Na3Bi_surfstate}(b),
the solid Dirac cone from bulk states is well separated from the topological
surface Dirac cone. While the surface states merge into the bulk states
at the Dirac points where bulk band gap closes. Thus, if Fermi level passes
through the bulk Dirac nodes, the surface Fermi surface has Fermi arc structures
as shown in Fig.~\ref{fig:Na3Bi_surfstate}(c).  

It seems that the entire fermi surface is closed, while its derivative and Fermi 
velocity are ill defined at the two singular points (corresponding to the projection 
of bulk Dirac points to the surface). The spin texture of surface states has helical
structure, quite similar to that of topological insulators, but the magnitude
of spin vector vanishes at the singular points. This kind of Fermi surfaces
has never been found before, and it can be understood following the
discussions for Weyl semimetal.~\cite{WanXG_WeylTI_2011, XuGang_HgCrSe_2011_PRL} 
If the 4$\times$4 Dirac point is split into two separated 2$\times$2 Weyl points in
momentum space by breaking time reversal or inversion
symmetry~\cite{Burkov_Weyl_2011,Burkov_Topological_nodal_semimetals_2011PRB}, 
the Fermi surface of surface states will also split into open segments which are Fermi arcs
discussed in Weyl semimetal. In Fig.~\ref{fig:Na3Bi_surfstate}(d), an exchange field breaking
time-reversal symmetry separates the touch of two Fermi arcs.

All these characters in contrast to conventional metals and topological insulators have been
be experimentally confirmed by successive ARPES measurements.~\cite{ChenYL_Na3Bi_2014Science, Xu2013, xu_observation_2014} 
The extraordinary magneto transport property in DSM Na$_3$Bi due to chiral anomaly 
has also been confirmed experimentally.~\cite{Ong_Na3Bi_2015, Xiong_Science_2015, Burkov_Science_2015}

{\bf 2.2: Cd$_3$As$_2$}
\hspace*{2pt}


After the theoretical prediction of DMS in Na$_3$Bi, the experimental
works to confirm it have been intensively performed. It is soon found to be unstable in air. 
Therefore, the present authors and their collaborators were motivated to find other better
materials to be more suitable for experimental studies. The well known compound 
Cd$_3$As$_2$ was re-investigated and was found to be another DSM candidate in 2013.~\cite{WangZJ_Cd3As2_2013}
Similarly, it has a single pair of Dirac points in bulk and Fermi arcs on proper surface. 
In particular, in one of its phase the spin degeneracy of crossing bands is lifted away from 
the Dirac points, in star contrast to other examples~\cite{young_dirac_2012, WangZJ_Na3Bi_2012}. 
It is further suggested that its quantum well structure can naturally 
support the quantum spin Hall (QSH) effect. The other nice aspect of Cd$_3$As$_2$ is 
its high carrier mobility up to 15000 cm$^2$V$^{-1}$s$^{-1}$ 
at room temperature and 80000 cm$^2$V$^{-1}$s$^{-1}$ at 4 K, reported about 50 years
ago~\cite{mobility}. These advantages make it a promising 
candidate for future experimental studies and potential applications.

\begin{figure}[tbp]
\includegraphics[clip,scale=0.5]{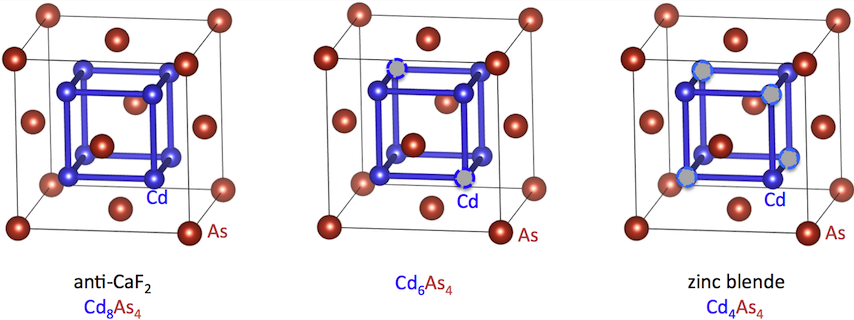}
\caption{(Color online) Crystal structure of Cd$_3$As$_2$. (left) The hypothetical aniti-CaF$_2$ structure 
of Cd$_8$As$_4$. (middle) Two diagonal sites with Cd vacancy with Cd$_6$As$_4$ formula as the 
building block of real Cd$_3$As$_2$ crystal structure. (right) The hypothetical zinc blende structure of Cd$_4$As$_4$. }
\label{fig:Cd3As2_CrystalStru}
\end{figure}

\begin{figure}[tbp]
\includegraphics[clip,scale=0.6]{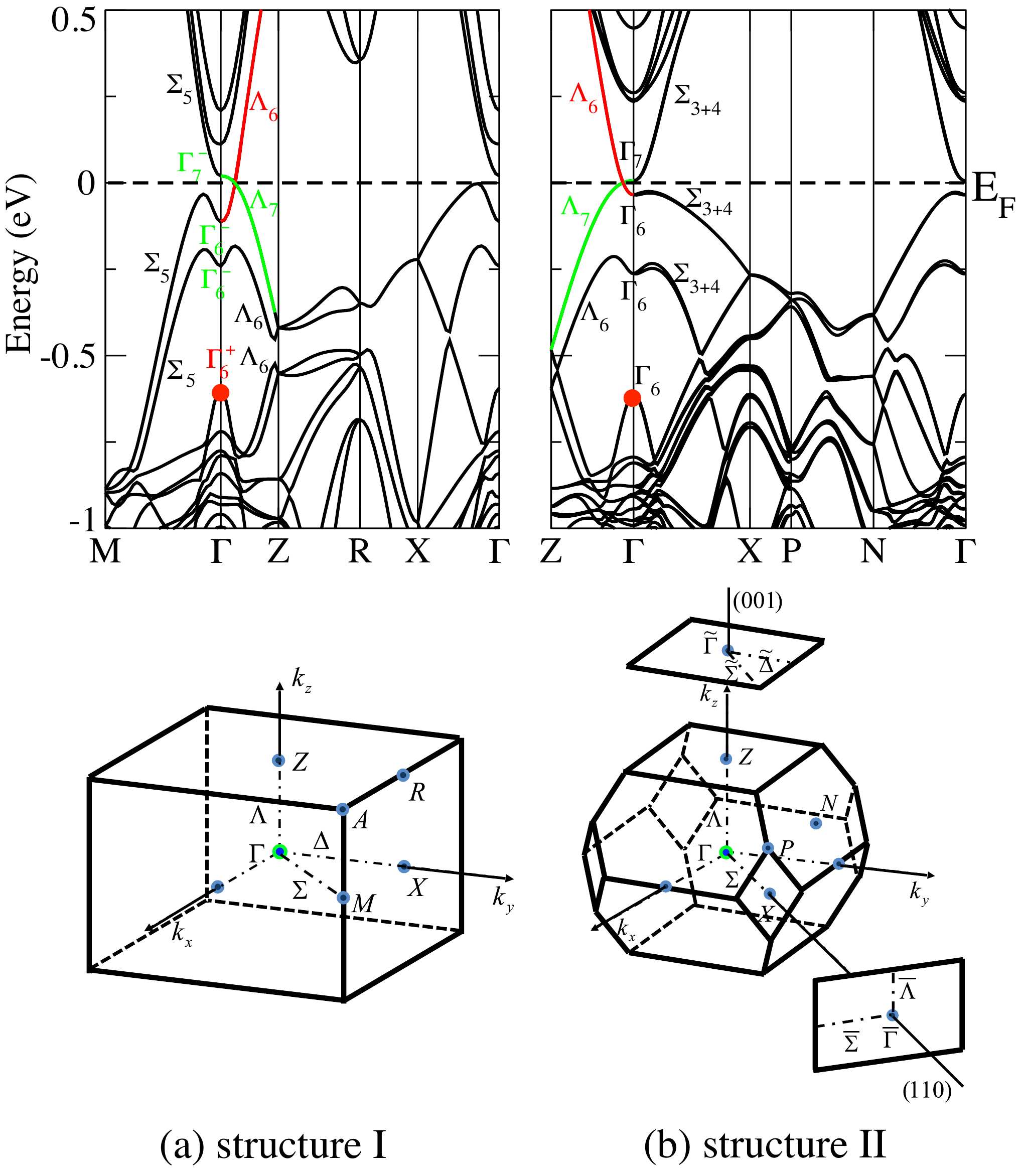}
\caption{(Color online) The calculated band structures for
  Cd$_3$As$_2$ with crystal structure I (a) and II (b), as well as the
  definition of high symmetric points in each Brillouin zone (BZ).
  The projected surface BZ for structure II is also shown. The
  representation of selected bands at $\Gamma$ and along high
  symmetric $\vec{k}$ path are indicated. Solid dots in band
  structures indicate the projected $s$ bands. }
\label{fig:Cd3As2_band}
\end{figure}

The crystal structure of Cd$_3$As$_2$ is complicated. In Fig.~\ref{fig:Cd3As2_CrystalStru},
there are hypothetical anti-fluorite Cd$_8$As$_4$ and zinc blende Cd$_4$As$_4$. 
The Cd$_6$As$_4$ is obtained with two diagonal sites of Cd vacancy in Cd$_8$As$_4$. This
is the basic building block of real Cd$_3$As$_2$ crystal structure. If the arrangement of
building block is random, i.e., the distribution of these vacancies is random, one may treat 
it by virtual crystal approximation (VCA) for simplicity~\cite{vcafluorite,vcazb}. However, 
those vacancies are in fact ordered even at room temperature, leading to a tetragonal
structure with $D_{4h}^{15}$ ($P4_{2}/nmc$) symmetry with 40 atoms per
unit cell being $\sqrt{2}\times2\times2$ supercell of building block. (called Structure I hereafter ), 
or a body centered tetragonal structure with $C_{4v}^{12}$ ($I4_{1}cd$) symmetry with 80
atoms per unit cell being $2\times2\times4$ supercell of building block (called Structure II hereafter). 
The later one is more favored~\cite{Cd3As2Struct} and we noticed there is correction~\cite{Cd3As2_Cava2014} 
to the crystal structure after our original work published. This vacancy ordering and
very large uint cell of Cd$_3$As$_2$ has brings serious problems
for theoretical studies in history, and there has been no first-principles
calculation before Ref.~\onlinecite{WangZJ_Cd3As2_2013}. 

Cd$_3$As$_2$ belongs to the II$_3$-V$_2$-types narrow gap semiconductor family.~\cite{mobility} 
It has drawn a lot attention since it has been believed to have inverted
band structure,~\cite{invert_exp1, invert_exp2, invert_theory1, invert_theory2} whereas all others
Cd$_3$P$_2$, Zn$_3$As$_2$ and Zn$_3$P$_2$ have normal band
ordering. In contrast to other inverted band compounds (like HgTe,
HgSe, and $\alpha$-Sn), Cd$_3$As$_2$ belongs to tetragonal symmetry,
and is the representative of this group, which has the splitted valence band top at $\vec{k}$=0.  

Similar to most of the semiconductors with anti-fluorite or
zinc-blende structures, the low energy electronic properties of
Cd$_3$As$_2$ are mostly determined by the Cd-$5s$ states (conduction
bands) and the As-$4p$ states (valence bands), as shown in
Fig.~\ref{fig:Cd3As2_band}. However, there are two distinct features: (1) band-inversion
around $\Gamma$ with the $s$-band (red solid cycle) lower than
the $p$ ones, which is an important sign of non-trivial topology;
(2) it is semimetallic with band crossings along the $\Gamma$-Z path.
This band-crossing is unavoidable since the two bands
belong to different ($\Lambda_6$ and $\Lambda_7$) representations
respectively, as distinguished by $C_{4}$ rotational symmetry around
$k_z$ axis. The different representation prohibits hybridization
between them, resulting in the protected band-crossing. The crossing points locate 
exactly at the Fermi level due to charge neutrality requirement and the Fermi surface 
has only a single pair of Dirac points (two symmetric points along $\Gamma$-Z related by TR).  
Structure I and II share these common features with small but remarkably difference as
will be addressed in the following.

As had been attempted with perturbation method~\cite{vacant}, vacancy
ordering and BZ folding play important role for the band-inversion, in
contrast to the cases of HgTe or Ag$_2$Te~\cite{Ag2Te}, where it was
driven by the the shallow $d$ states. To prove this, Cd$_3$As$_2$ with 
hypothetic anti-fluorite structure without vacancy in Fig.~\ref{fig:Cd3As2_CrystalStru}(a) has
been calculated by using VCA method, compared with the usual first-principles
calculation for Cd$_3$As$_2$ in the same anti-flurite structure but with
Cd vacancy in Fig.~\ref{fig:Cd3As2_CrystalStru}(b). It is found that the
VCA calculation gives out normal band ordering at $\Gamma$ but the one with
Cd vacancy results in inverted band ordering. At the
BZ boundary X point of the hypothetic anti-fluorite structure without
Cd vacancy, there exists shallow $s$ and $p$ states, which will be folded 
to $\Gamma$ in realistic structure with Cd vacancy ordering. Therefore, the
hybridization among the states with the same representation will push
them away from each other, i.e, make the lowest $s$-state
even lower and highest $p$-state higher, resulting in the robust band
inversion at $\Gamma$. The band-inversion calculated from the
generalized gradient approximation (GGA) is about 0.7 eV for both
structure I and II. The possible underestimation of the $s$-$p$ gap within GGA
is improved by the calculations with HSE
method.~\cite{HSE03,HSE06} The band-inversion around 0.3 eV is obtained,
being consistent with most of the existing experimental evidence, such
as the optical and transport measurements.~\cite{invert_exp1, invert_exp2}

The inverted band structure discussed above suggests that Cd$_3$As$_2$
is topologically non-trivial. However, due to the four-fold rotational
symmetry, it is in fact a 3D DSM with a pair of 3D Dirac
points at the Fermi level. This is slight different from Na$_3$Bi. Similarly,
if the four-fold rotational symmetry is broken, two Weyl fermions with opposite
chirality will be annihilate each other, resulting in a massive Dirac fermions 
with finite band gap. The resulting insulator is a strong TI.

\begin{figure}[tbp]
\includegraphics[clip,scale=0.45,angle=0]{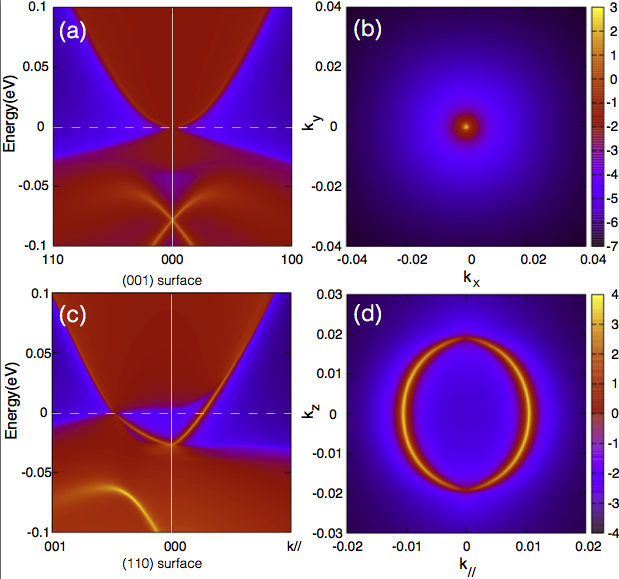}
\caption{(Color online) The calculated surface states (left panels)
  and corresponding fermi surface (right panels) of structure II
  Cd$_3$As$_2$ for its (001) (upper panels) and (110) (lower panels)
  surface.}
\label{fig:Cd3As2_surfState}  
\end{figure}

Similar as Na$_3$Bi, Cd$_3$As$_2$ has coexistence of bulk and surface 
Dirac cone in its surface state. The semi-infinite (001) and (110) surfaces of
the structure II Cd$_3$As$_2$ are presented in Fig.~\ref{fig:Cd3As2_surfState}.  
For (001) surface, the surface projection of continuous bulk states superposes the
non-trivial surface states. The Fermi surface of this surface is just
a point as shown in Fig.~\ref{fig:Cd3As2_surfState}(b). For (110) surface, 
the non-trivial surface states are clear. The Fermi surface of
such topological surface states has two half-circle Fermi arcs touching
at the singularity points where the surface projection of bulk Dirac points exists.

\begin{figure}[tbp]
\includegraphics[clip,scale=0.7,angle=0]{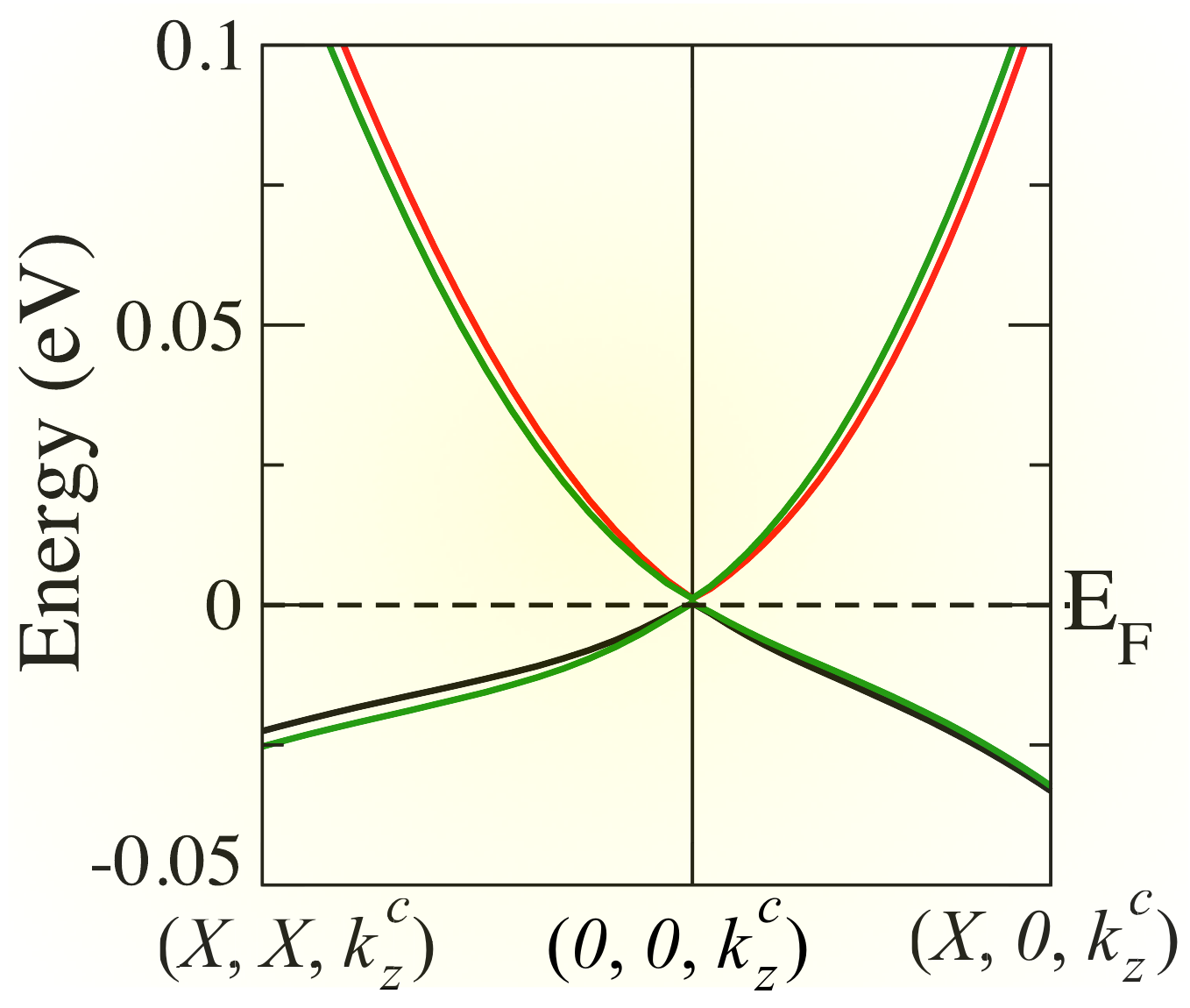}
  \caption{(Color online) Band dispersions and band-splitting
    in the plane passing through Dirac point (0,0,$k_z^c$) and perpendicular 
    to $\Gamma$-Z for structure II. $X$ and $k_z^{c}$ are around 0.1 and 0.032 \AA$^{-1}$, respectively.}
    \label{fig:Cd3As2_spinsplit}  
\end{figure}

It is noted that structure II has no inversion symmetry and no Kramer degeneracy
at general momenta. As shown in Fig.~\ref{fig:Cd3As2_spinsplit}, the bulk Dirac
cone in structure II has spin splitting as momentum deviate from the rotation axis
due to the inversion symmetry broken. Such spin polarized bulk Dirac cone is quite
similar to the Weyl cone in WSM of TaAs family.

Soon after this theoretical prediction, the present authors and their collaborators
have observed the 3D bulk Dirac cone~\cite{ChenYL_Cd3As2_2014NatMa, PhysRevLett.113.027603} and surface states~\cite{ZhouXJ_Cd3As2_2014} 
in Cd$_3$As$_3$ through ARPES measurement. Other experimental works
include Ref.~\onlinecite{neupane_observation_2013, jeon_landau_2014, PhysRevLett.113.246402, Liang2015, Feng2015}.

B. Yang and N. Nagaosa have classified Na$_3$Bi and Cd$_3$As$_2$ as DSM
carrying quantized topological invariant,~\cite{PhysRevB.92.165120} while BiO$_2$~\cite{young_dirac_2012} 
and distorted spinel~\cite{PhysRevLett.112.036403} as the class having only
a single Dirac point at a time-reversal invariant momentum. The classification
of DSM is further investigated by Gibson {\it et al.}~\cite{PhysRevB.91.205128} and
several similar DSM candidates have also been proposed by Du {\it et al.}~\cite{Wan2015}
As originally pointed out by S. Murakami,~\cite{murakami_phase_2007, Murakami2011748}, 
DSM can also be obtained at the topological phase transition point for IS conserved system.
Bi$_{1-x}$Sb$_x$~\cite{ZhangHJ_BiSb_2009PRB, PhysRevLett.111.246603} is believed to be DSM at 
the phase transition point from TI to normal insulator. We have proposed that 
ZrTe$_5$ and HfTe$_5$ are DSM only at the phase transition point
from weak to strong TI.~\cite{ZrTe5_PRX_2014,Li2014,WanNLZrTe52015,WanNLZrTe5Opt2015}

{\large\bf 3. Weyl Semimetal}
\hspace*{2pt}


As discussed before, DSM has Dirac nodes composed by two opposite chiral Weyl nodes 
``kiss" at the same $k$-point due to the degeneracy in spin degree of freedom. In 
condensed matters, such spin degeneracy at general $k$-point is usually 
protected by coexistence of time-reversal symmetry (TRS) and inversion (IS). It is 
known as Kramer degeneracy. Breaking either TRS or IS might result 
in two-fold degenerate Weyl nodes. This leads to two types of WSM,
i.e., the magnetic WSM and nonmagnetic noncentrosymmetric WSM. The first WSM
was proposed to be in pyrochlore iridates with all-in/all-out magnetic ordering 
by X. Wan {\it et al.}~\cite{WanXG_WeylTI_2011} In the same year, our collaborators
and us~\cite{XuGang_HgCrSe_2011_PRL} proposed that ferromagnetic half-metal HgCr$_2$Se$_4$ might be
a WSM with pair of Weyl nodes having topological charge of 2, so called double WSM.~\cite{CFang_doubleWeyl} 
However, the magnetic WSM suffers from the complex domain structure 
in ARPES measurement. There is still no material definitely confirmed to be 
magnetic WSM up to now. Finding a nonmagnetic WSM is thus very 
important. In this section, we will introduce our theoretical predictions of
FM HgCr$_2$Se$_4$ and noncentrosymmetric nonmagnetic TaAs family.
The recent progress in nonmagnetic WSM is the proposal of type-II WSM~\cite{Soluyanov2015}
in WTe$_2$.~\cite{Ali2014hj}

{\bf 3.1: Magnetic Weyl Semimetal: HgCr$_2$Se$_4$}
\hspace*{2pt}


In 2011, it was proposed that a spinel structural compound HgCr$_2$Se$_4$ might 
be a WSM under its ground state. Compared with the former proposal of WSM~\cite{WanXG_WeylTI_2011}, 
HgCr$_2$Se$_4$ has only one 
pair of Weyl cones around Fermi level and each Weyl cone has topological charge of 2.
HgCr$_2$Se$_4$ has quite simple ferromagnetic (FM) ordering and robust against 
choice of on-site Coulmb interaction parameter $U$. Before WSM state was proposed 
in it, HgCr$_2$Se$_4$ was known as a FM spinel exhibiting large coupling effects
between electronic and magnetic properties~\cite{HgCrSe-rev}. It shows
quite interesting properties like giant magnetoresistance~\cite{HgCrSe-MR},
anomalous Hall effect~\cite{HgCrSe-AHE}, and red shift of optical
absorption edge~\cite{HgCrSe-OP}. Its experimental Curie temperature $T_c$ is high
(around 106$\sim$120 K), and the saturated moment is around 5.64
$\mu_B$/f.u.~\cite{HgCrSe-Tc,HgCrSe-M}, approaching the atomic value
expected for high-spin Cr$^{3+}$.  Its transport behavior is different
from other FM chalcogenide spinels, like CdCr$_2$Se$_4$ and
CdCr$_2$S$_4$ which are clearly semiconducting. HgCr$_2$Se$_4$
exhibits semiconducting character in the paramagnetic state but
metallic in the low temperature FM
phase~\cite{HgCrSe-MR,HgCrSe-Tran-1,HgCrSe-Tran}.  

\begin{figure}[tbp]
\includegraphics[clip,scale=0.45]{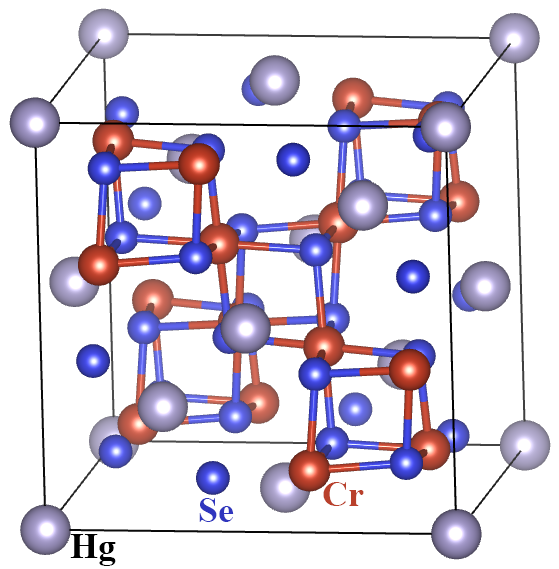}
\caption{(Color online) Crystal structure of spinel HgCr$_2$Se$_4$. }
\label{fig:HgCr2Se4_CrystalStru}
\end{figure}

The spinel structure, with space group Fd$\bar{3}$m, can be related to the zinc blende
and/or diamond structures. As shown in Fig.~\ref{fig:HgCr2Se4_CrystalStru}, taking Cr$_2$Se$_4$ 
cluster as a single pseudo-atom (called X) located at the center of mass, HgX becomes 
a typical zinc blende structure. The pseudo atom X is actually a small cube
formed by 4 Cr atoms and 4 Se atoms located at the cube corners
alternatively. The cubes are connected by sharing the Cr atoms at the
corner. As the result, each Cr atom is octahedrally coordinated by 6
nearest Se atoms. There are two HgX formula unit (f.u.) in each unit
cell, and they are connected by the inversion symmetry similar to the two atoms in
the diamond structure. 

\begin{figure}[tbp]
\includegraphics[clip,scale=0.5]{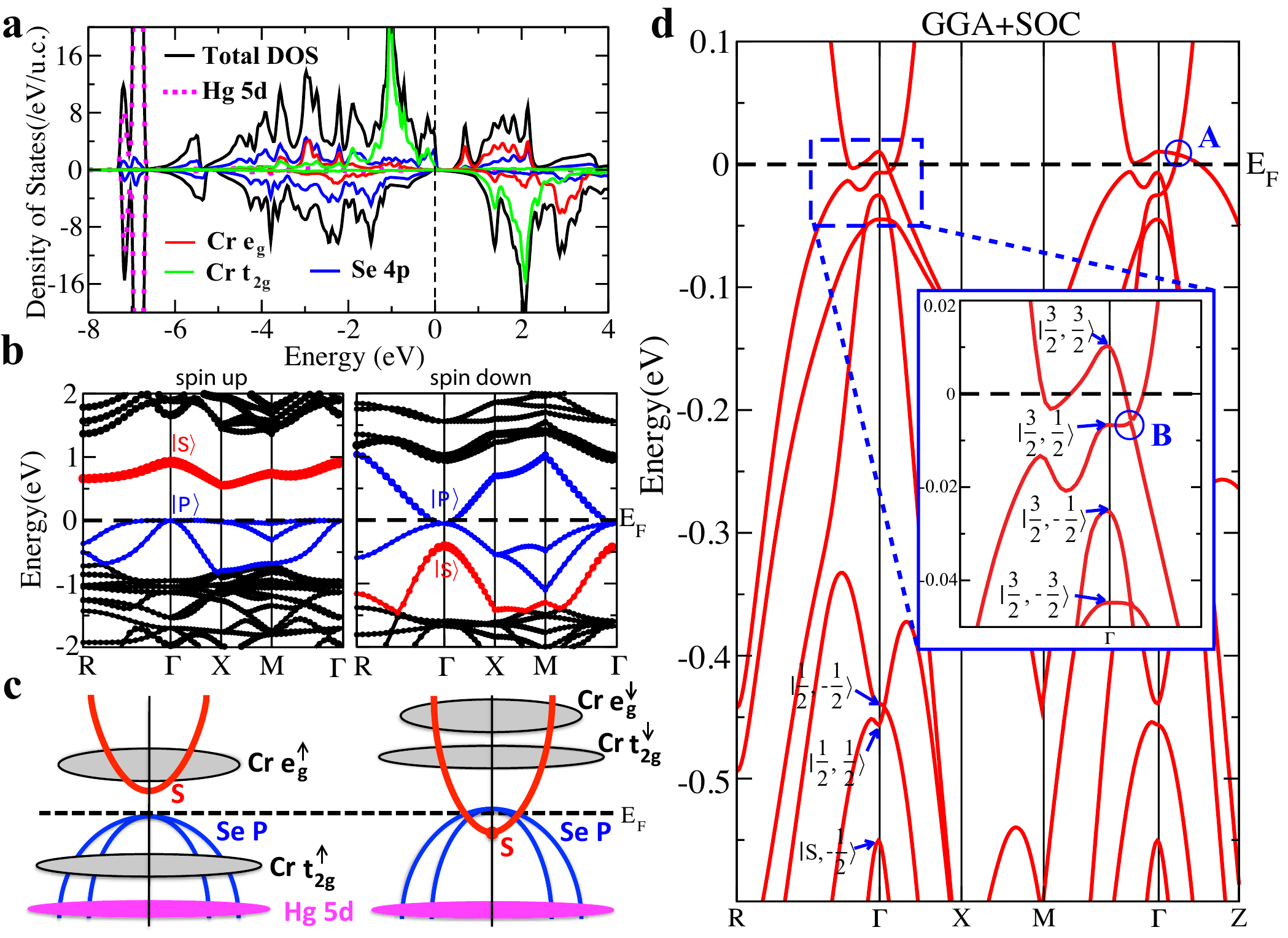}
\caption{(Color online) {\bf Electronic Structures of HgCr$_2$Se$_4$.}
  (a) The total and partial density of states (DOS); (b) The band
  structures without SOC (showing the up and down spin parts
  separately); (c) The schematic understanding for the band-inversion,
  where the $|S\rangle$ state is lower than the $|P\rangle$ states in
  the down spin channel; (d) The band structure after including SOC
  (with majority spin aligning to the (001) direction). The low energy
  states at $\Gamma$ are indicated as explained in the main text. }
  \label{fig:HgCr2Se4_band}
\end{figure}

The first-principles calculation confirms that FM solution is considerably
more stable than non-magnetic solution by about 2.8eV/f.u. lower in totoal
energy. The obtained magnetic moment (6.0 $\mu_B$/f.u.) agrees with
experimental value~\cite{HgCrSe-Tc,HgCrSe-M} very well. The electronic structures
shown in Fig.~\ref{fig:HgCr2Se4_band} (a) and (b) suggest that it can be
approximately characterized as a ``zero-gap half-metal'' in the case
without spin-orbit coupling (SOC). It is zero-gap because there is
band touching around $\Gamma$ at Fermi level in spin down 
channel; it is half-metal since there band gap in the up-spin channel. 
The $3d$-states of Cr$^{3+}$ are in the configuration of
$t_{2g}^{3\uparrow}e_g^{0\uparrow}t_{2g}^{0\downarrow}e_g^{0\downarrow}$.
The band gap is due to crystal field splitting of the $t_{2g}^{3\uparrow}$ and
$e_g^{0\uparrow}$ manifolds. The Se-$4p$ states are from -6 to 0 eV, being
fully occupied and contribute to the valence band maxima dominantly. 
By hybridizing with Cr-$3d$ states, Se-$4p$ are slightly 
spin-polarized but with opposite moment (about -0.08 $\mu_B$/Se) to that on Cr. 
The zero-gap behavior in spin down channel is the most important 
character. It suggests the inverted band structure 
around $\Gamma$, which is similar to the case in HgSe or HgTe~\cite{HgTe-1,HgTe-2}.

The four low energy states (eight after considering spin degree of freedom) at
$\Gamma$ can be labelled as $|P_x\rangle$, $|P_y\rangle$,
$|P_z\rangle$, and $|S\rangle$, which are linear combinations of
atomic orbitals:
$|P_{\alpha}\rangle\approx\frac{1}{\sqrt{8}}\sum_{i=1}^{8}|p_{\alpha}^{i}\rangle$,
and $|S\rangle\approx 0.4\sum_{j=1}^{2}|s^{j}\rangle+0.24
\sum_{k=1}^{4} |d_{t_{2g}}^{k}\rangle$, where $\alpha=x,y,z$ and
$i,j,k$ runs over Se, Hg, Cr atoms in the unit cell;
$|s\rangle$, $|p_{\alpha=x,y,z}\rangle$,
$|d_{t_{2g}=xy,yz,zx}\rangle$ are the atomic orbitals of
each atom. Taking these four states as basis, it is found to be
the same situation in HgSe or HgTe. The only difference is the
presence of exchange splitting in HgCr$_2$Se$_4$. The band inversion,
$|S,\downarrow\rangle$ being lower than $|P,\downarrow\rangle$ at $\Gamma$
as shown in Fig.~\ref{fig:HgCr2Se4_band}(c), is due to 
the following two factors. Firstly,
the Hg-$5d$ states are very shallow (located at about -7.0 eV in 
Fig.~\ref{fig:HgCr2Se4_band}(a)) and its hybridization with Se-$4p$ 
states pushes the antibonding Se-$4p$ states higher, similar to 
that in HgSe. Secondly, the hybridization between 
unoccupied Cr-$3d^\downarrow$ and
Hg-$6s^\downarrow$ states in the down spin channel pushes the
Hg-$6s^\downarrow$ state lower in energy (Fig.~\ref{fig:HgCr2Se4_band} (b) and (c)). 
Therefore, the $|S,\downarrow\rangle$ is about 0.4eV lower than
$|P,\downarrow\rangle$ states at $\Gamma$, and it is further enhanced to be
0.55eV in the presence of SOC. It is important to check the
correlation effect beyond GGA. It has been shown that semiconducting
CdCr$_2$S$_4$ and CdCr$_2$Se$_4$ can be well described by the
LDA+$U$ calculations with effective $U$ around $3.0$
eV~\cite{CdCrS-ldau,ACrX}. The same LDA+$U$
calculations for HgCr$_2$Se$_4$ show that the band-inversion
remains unless the $U$ is unreasonably large ($>8.0$ eV).

\begin{figure}[tbp]
\includegraphics[clip,scale=0.5]{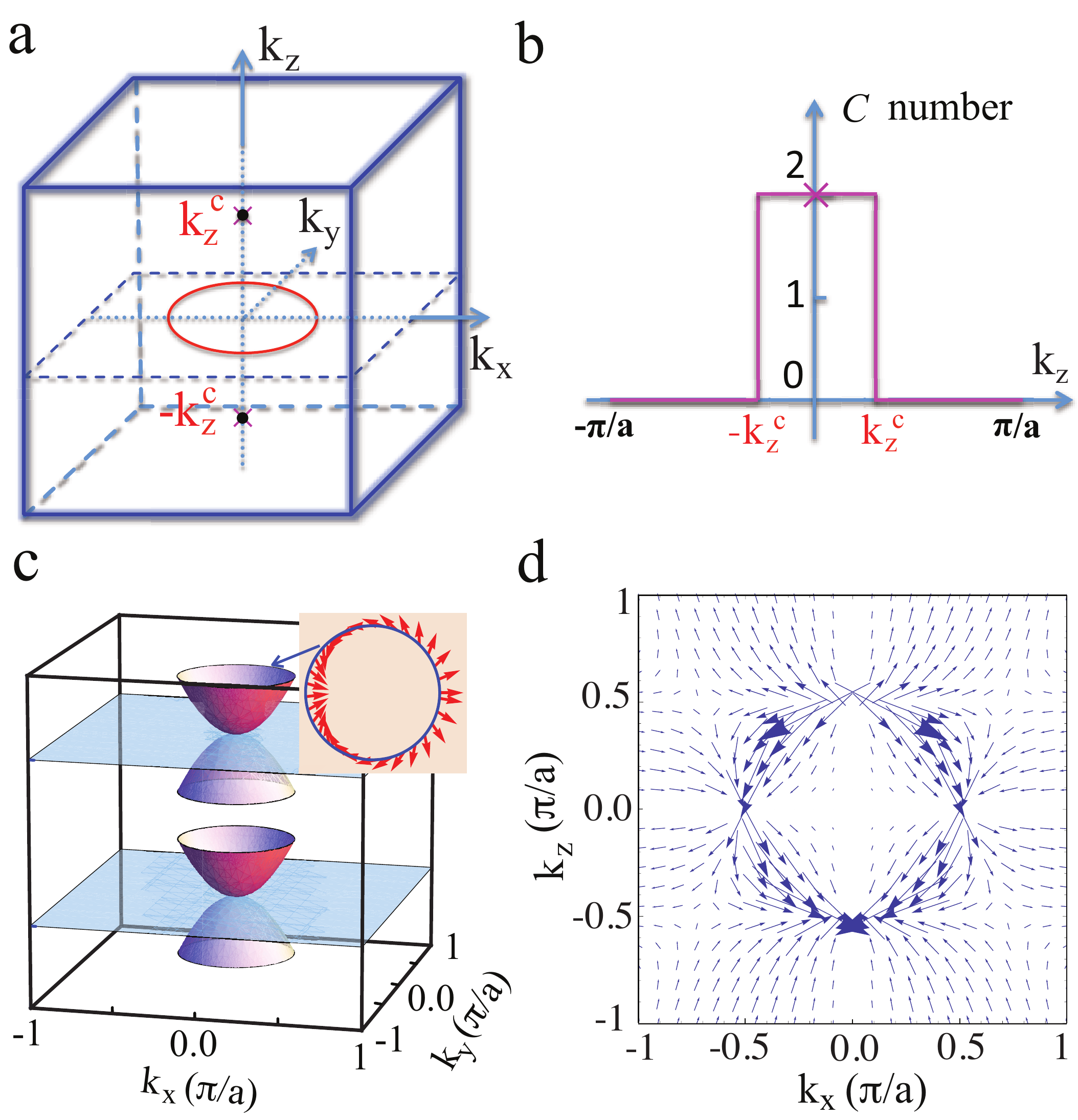}
\caption{(Color online) {\bf Weyl nodes and Berry flux in momentum space of
    HgCr$_2$Se$_4$}. (a) The band-crossing points $\pm k_z^c$; (b) The 
    Chern number of occupied states in 2D planes perpendicular to $k_z$ axis; 
    (c) The schematic plot of the band dispersion around the Weyl nodes in the
  $k_z$=$\pm k_z^c$ plane, and the inset shows the chiral spin
  texture. (d) The Berry flux evaluated as Berry's curvature in the
  ($k_x$, $k_z$) plane with $k_y$=0.}
\label{fig:HgCr2Se4_chernNum}
\end{figure}

Considering SOC, the new low energy eigen states at $\Gamma$
are given as $|\frac{3}{2},\pm\frac{3}{2}\rangle$,
$|\frac{3}{2},\pm\frac{1}{2}\rangle$,
$|\frac{1}{2},\pm\frac{1}{2}\rangle$, and
$|S,\pm\frac{1}{2}\rangle$, which can be constructed from the
$|P\rangle$ and $|S\rangle$ states. Using
this effective hamiltonian, the low energy band structures of
HgCr$_2$Se$_4$ can be well reproduced. Due to the exchange
splitting, the eight states at $\Gamma$ are all
energetically separated, with the $|\frac{3}{2},\frac{3}{2}\rangle$
having the highest energy, and the $|S,-\frac{1}{2}\rangle$ being
the lowest. Several band-crossings are
observed as shown in the band structure of Fig.~\ref{fig:HgCr2Se4_band}(d). 
Among them, two kinds of band-crossings (called A and B) are
important for the states very close to the Fermi level. The
crossing-A gives two points located at $k_z=\pm k_z^c$ along the
$\Gamma -Z$ line, and the trajectory of crossing-B is a closed loop
surrounding $\Gamma$ point in the $k_z$=0 plane, as schematically
shown in Fig.~\ref{fig:HgCr2Se4_chernNum}(a).  For the 2D planes 
with fixed-$k_z$ ($k_z\ne 0$ and $k_z\ne\pm k_z^c$), the band 
structures are all gapped in the sense that a curved Fermi level is 
defined. Thus, the Chern number $C$ for 
each $k_z$-fixed plane can be evaluated. It is found
that $C=0$ for the planes with $k_z<-k_z^c$ or $k_z>k_z^c$, while
$C=2$ for the planes with $-k_z^c<k_z<k_z^c$ and $k_z\ne 0$.  The 
crossing-A points locate at the phase boundary
between $C=2$ and $C=0$ planes, i.e. $k_z=\pm k_z^c$ plane, and they 
are topologically unavoidable Weyl nodes.
On the other hand, however, the crossing-B points forming the closed loop in
$k_z=0$ plane are just accidental and it is due to the presence of
crystal mirror symmetry with respect to the $k_z$=0 plane. The
crossing-B is not as stable as crossing-A in the sense that it can
be changed by changing of the crystal symmetry. This node-line becomes
a hot topic recently as we will discuss in the next section.

\begin{figure}[tbp]
\includegraphics[clip,scale=0.40]{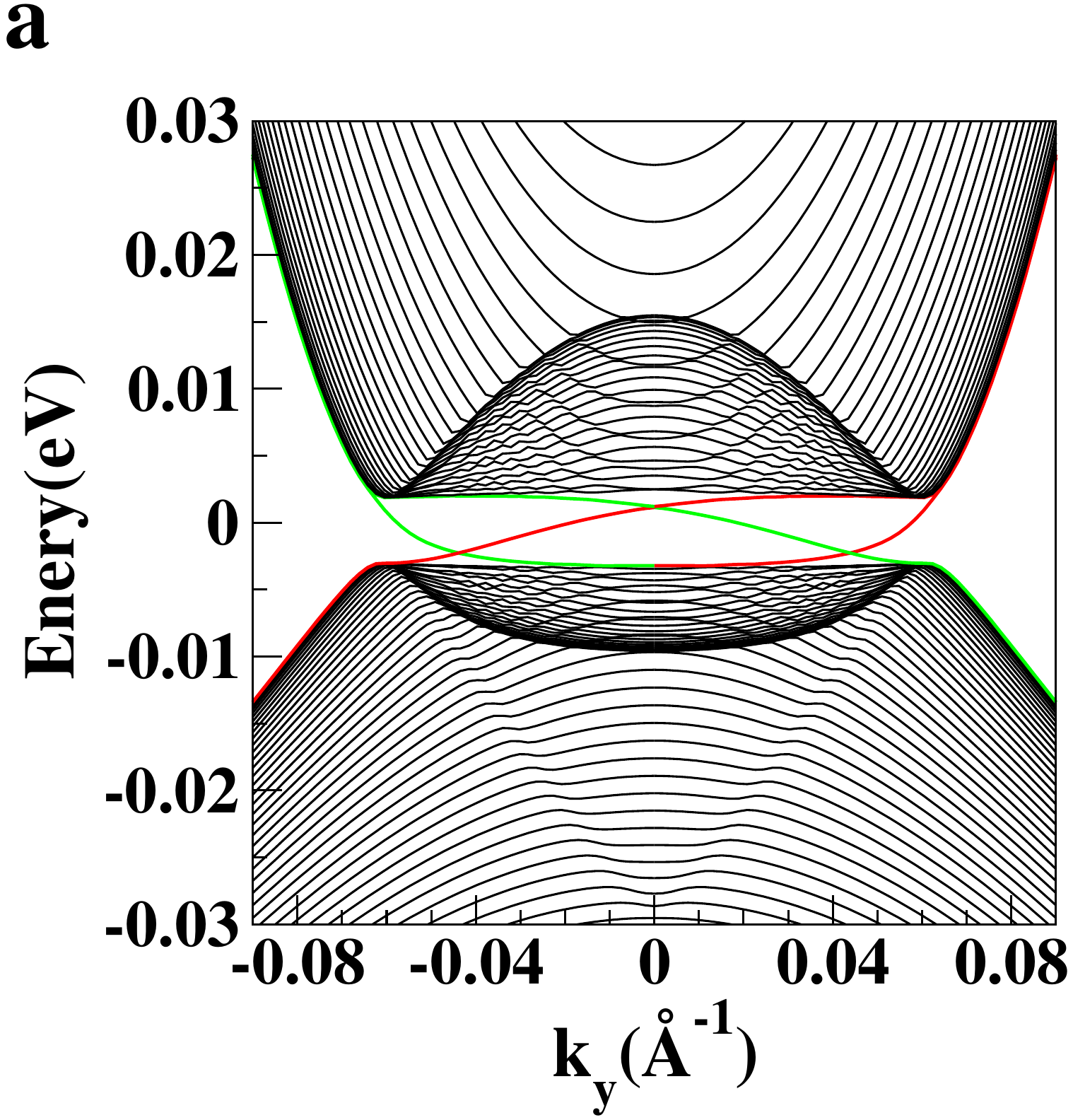}
\includegraphics[clip,scale=0.25]{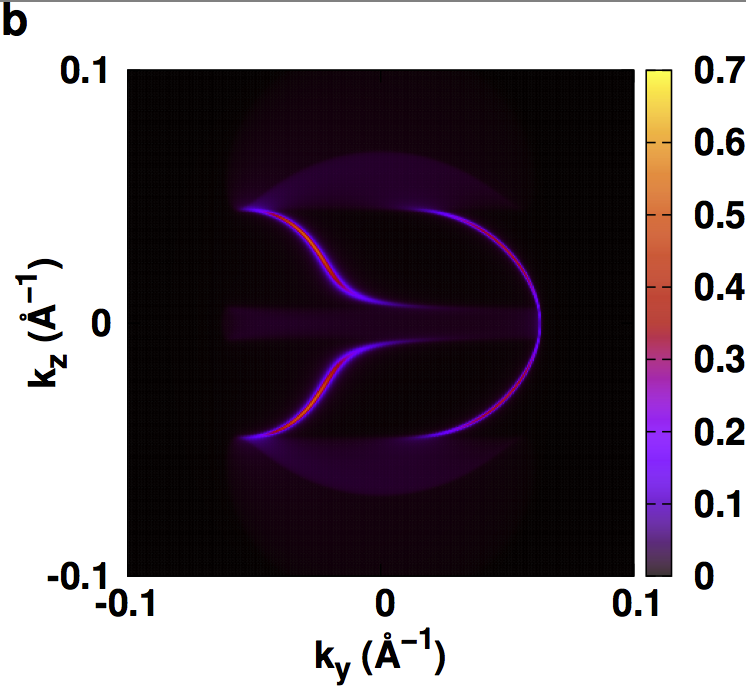}
\caption{(Color online) {\bf Edge states and Fermi arcs of
    HgCr$_2$Se$_4$.}  (a) The calculated edge states for the plane
  with $k_z$=0.06$\pi$. A ribbon with two edges is used, and there are
  two edge states for each edge (because $C$=2). The states located at
  different edges are indicated by different colors. (b) The
  calculated Fermi arcs for the ($k_y$,$k_z$) side surface.}
\label{fig:HgCr2Se4_arc}  
\end{figure}

Considering the minimum two basis $|\frac{3}{2},\frac{3}{2}\rangle$
and $|S,-\frac{1}{2}\rangle$ which catch the band-inversion nature,
the 8$\times$8 Hamiltonian can be downfolded into an effective
2$\times$2 model by only keeping these two states and integrating
out all the other bands, which reads
\begin{equation}
H_{eff}=\left[
\begin{array}{cc}
M &  Dk_zk_-^2  \\
 Dk_zk_+^2 & -M
\end{array}
\right].
\end{equation}
Here $k_\pm=k_x\pm i k_y$, and $M=M_0-\beta k^2$ is the mass term
up to the second order. $M_0>0$ and $\beta>0$ are the condition
to ensure the band inversion. Since the two basis have opposite parity,
the off-diagonal element has to be odd in $k$. The
$k_\pm^2$ term is necessary to conserve the angular moment along
$z$-direction. Therefore, to the leading order, the $k_zk_\pm^2$ is
the only possible form for the off-diagonal element. This two level hamiltonian
has eigen values $E(k)=\pm\sqrt{M^2+D^2k_z^2(k_x^2+k_y^2)^2}$. 
Two gapless solutions: one is the degenerate points along the
$\Gamma -Z$ line with $k_z=\pm k_z^c=\pm\sqrt{M_0/\beta}$; the other
is a circle around the $\Gamma$ point in the $k_z=0$ plane
determined from the equation $k_x^2+k_y^2=M_0/\beta$. These are
consistent with the first-principles
calculations. The presence of $k_\pm^2$ in the off-diagonal
term~\cite{Onoda-2} makes the Chern number $C$ being 2 
for the planes of $-k_z^c<k_z<k_z^c$ except the $k_z=0$ 
plane where node-line exists.
The in-plane band dispersions near the Weyl nodes at $k_z=\pm k_z^c$
are quadratic rather than linear, with a phase of 4$\pi$ for the chiral 
spin texture as shown in Fig.~\ref{fig:HgCr2Se4_chernNum}(c).
The two Weyl nodes located at $\pm k_z^c$ have opposite chirality, 
and they form a single pair
of magnetic monopoles carrying gauge flux in $\vec{k}$-space as
shown in Fig.~\ref{fig:HgCr2Se4_chernNum}(d). The band-crossing 
loop in the $k_z=0$ plane is not topologically unavoidable, however, its existence requires that
all gauge flux in the $k_z$=0 plane (except the loop itself) must be
zero.

This Chern semi-metal state realized in HgCr$_2$Se$_4$ will lead to
novel physical consequences, which can be measured experimentally.
The 8-bands Kane model fitted from first-principles calculations 
is used for the numerical demonstration. First, each $k_z$-fixed 
plane with non-zero Chern number can be regarded as a 2D Chern 
insulator, and there must be chiral edge states for such plane if an edge is created. By
Replacing $k_x$ with $-i\hbar\partial_x$ and using open boundary
condition along x (or y) direction, one can obtain the edge states
for each fixed-$k_z$. The number of edge states is two for the case
of $C$=2 as shown in Fig.~\ref{fig:HgCr2Se4_arc}(a), or zero for 
the case of $C$=0. If the
chemical potential is located within the gap, only the chiral edge
states can contribute to the Fermi surface, which are isolated
points for each Chern insulating plane but nothing for the plane
with $C$=0. Therefore the trajectory of such points in the ($k_x$,
$k_z$) surface or ($k_y$, $k_z$) surface form non-closed fermi arcs, 
which can be measured by ARPES as shown in Fig.~\ref{fig:HgCr2Se4_arc} (b).  
This is very much different from the conventional metals, where 
the Fermi surfaces must be either closed or interrupted by 
the Brillouin zone boundary. The possible Fermi arcs has been 
recently discussed from a view point of accidental degeneracy 
for pyrochlore iridates~\cite{WanXG_WeylTI_2011}. In
general, any magnetic system with band-crossing nodes at Fermi level may
show Fermi arcs on its surface. As demonstrate in Fig.~\ref{fig:HgCr2Se4_arc}, 
the calculated Fermi arcs for HgCr$_2$Se$_4$ not only 
end at $k_z$=$\pm k_z^c$, but also are interrupted by 
the $k_z$=0 plane, where the Node-line exists. Nevertheless, for WSM
the Fermi arcs should be stable because the 
Weyl nodes are topologically unavoidable.

\begin{figure}[tbp]
\includegraphics[clip,scale=0.4]{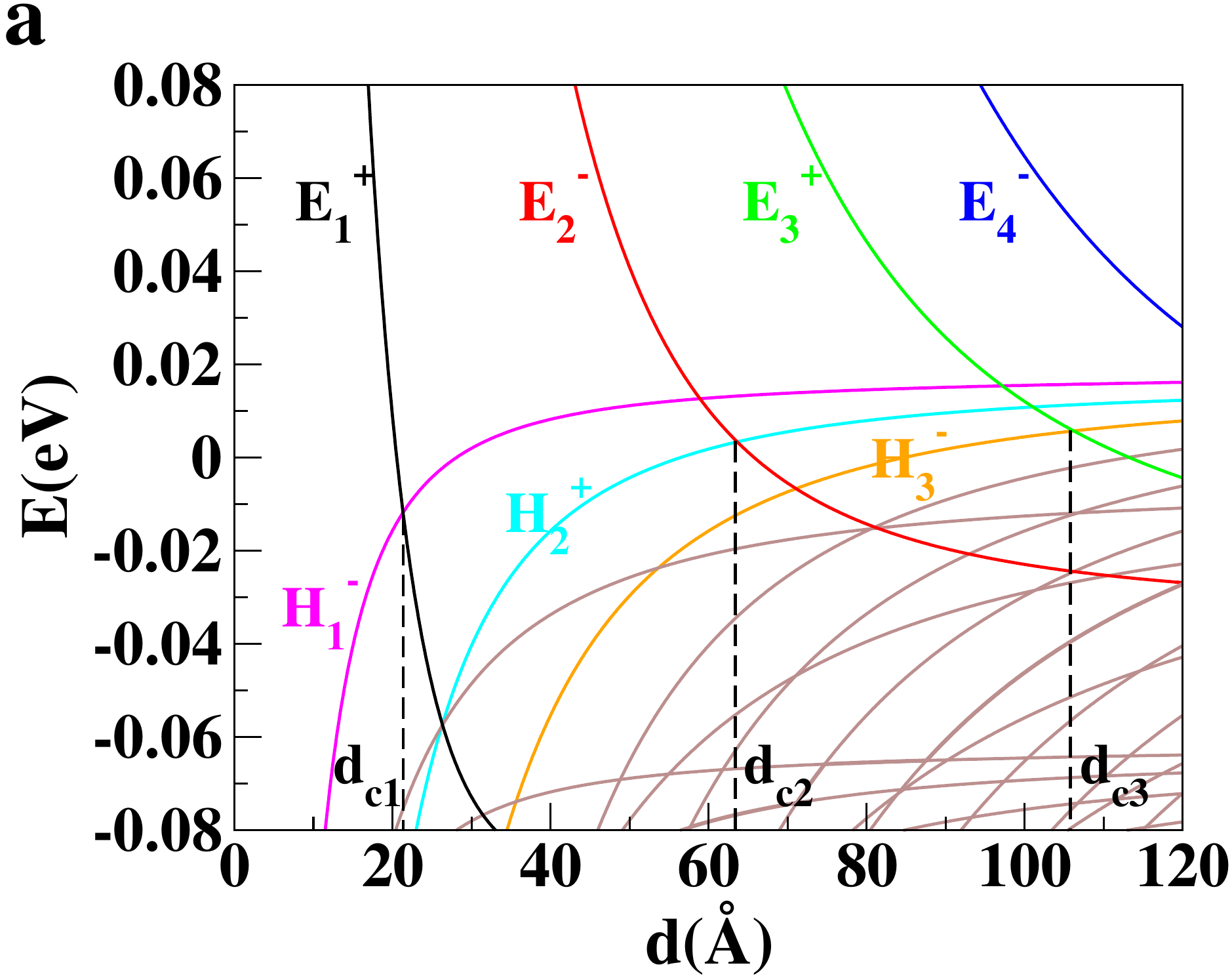}
\includegraphics[clip,scale=0.4]{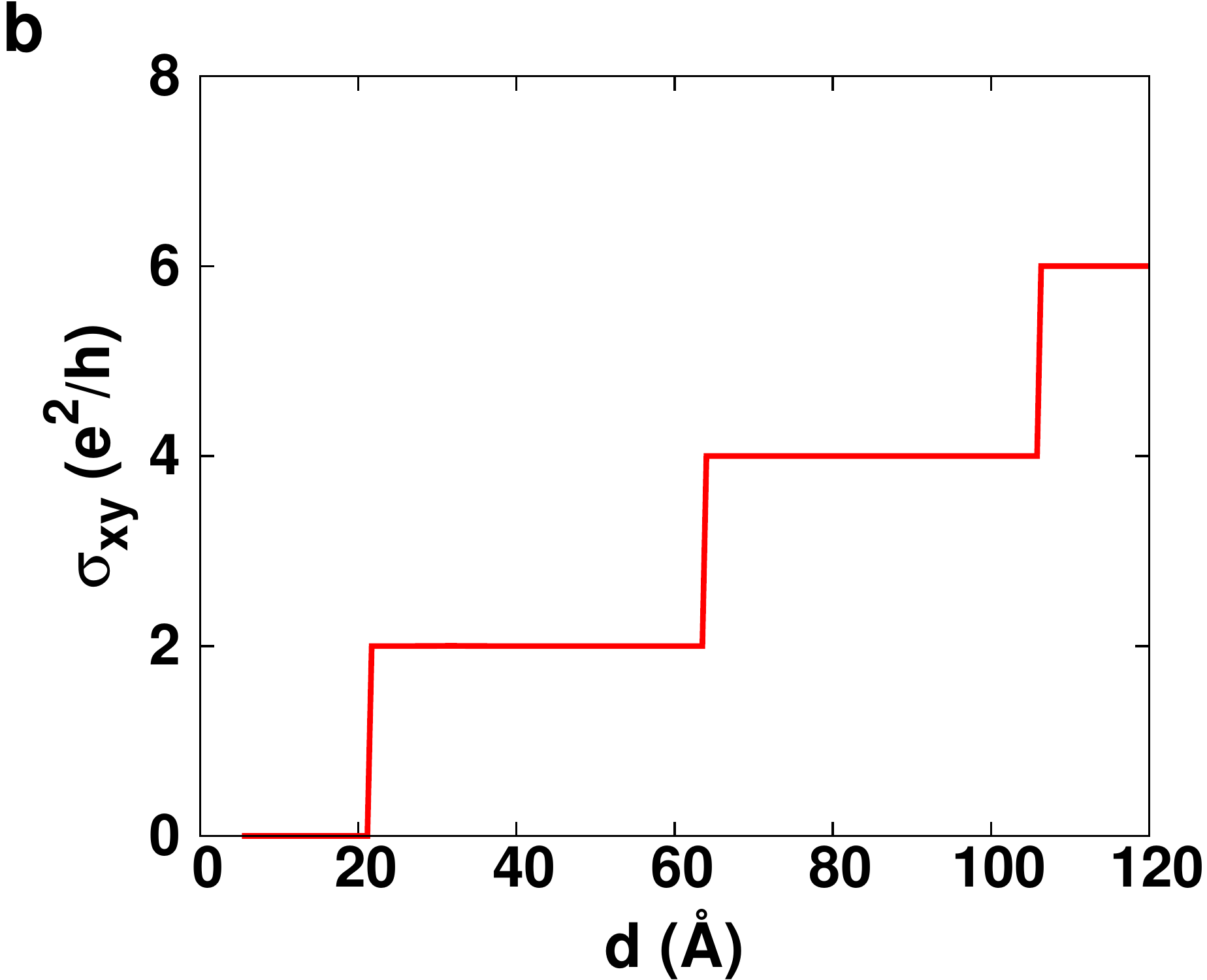}
\caption{(Color online) {\bf Quantized Anomalous Hall effect in
    HgCr$_2$Se$_4$ thin film.}  (a) The subband energy levels at
  $\Gamma$ point as function of film thickness. (b) The Hall
  conductance as function of film thickness.}
  \label{fig:HgCr2Se4_QAHE}  
\end{figure}

The QAHE, however, is an unique physical consequence characterizing
the WSM nature in ferromagnetic HgCr$_2$Se$_4$
quantum-well structure.~\cite{qahe_advphy} For 2D band insulators, the transverse Hall
conductance should be quantized as $\sigma_{xy}=C\frac{e^2}{h}$,
where $C$ is the Chern number.  Such a quantum Hall effect without
magnetic field has been long-pursued,~\cite{haldane_model_1988,onoda_quantized_2003,yu_quantized_2010, qahe_advphy} 
but only achieved experimentally in 2013 in magnetic ion doped topological insulator films
as theoretically predicted in 2010.~\cite{chang_experimental_2013, yu_quantized_2010, qahe_advphy} 
In HgCr$_2$Se$_4$, considering the $k_z$-fixed
planes, the Chern number $C$ is non-zero for limited regions of
$k_z$ due to the band inversion around $\Gamma$ as
discussed above. In the quantum well structure, the low
energy states around $\Gamma$ are quantized into
subbands, with $|H_n\rangle$ and $|E_n\rangle$ for hole and
electron subbands, respectively. All the subbands' energy level change as
function of film thickness. As shown in Fig.~\ref{fig:HgCr2Se4_QAHE}, when
the thickness of the film is very thin the band inversion in the
bulk band structure will be removed entirely by the finite size
effect. With the increment of the thickness, the band inversion 
among these subbands restores, which results in jumps in the Chern number or
the Hall coefficient $\sigma_{xy}$~\cite{HgMnTe, qahe_advphy}. As shown in
Fig.~\ref{fig:HgCr2Se4_QAHE}(b), the critical thickness of the film is around 21\AA.
The subsequent jumps of $\sigma_{xy}$ is in unit of $2e^2/h$ due to the change 
of 2 in Chern number. This brings an interesting way to realize high plateau quantum
Hall conductivity of 2 in QAHE, being distinguished from present experimentally achievable
magnetic doped TI system.~\cite{qahe_advphy}
In fact, the strong anomalous Hall effect has been observed for the
bulk samples of HgCr$_2$Se$_4$~\cite{HgCrSe-AHE}. This is in sharp
contrast to pyrochlore iridates, where the anomalous Hall effect
should be vanishing due to the AF ordering.

Inspired by this theoretical proposal, recent detailed transport studies have revealed that
the ground state of HgCr$_2$Se$_4$ has nearly full spin-polarized current in its $s$-orbital
conduction band, which is a strong evidence for its half-metallicity.~\cite{PhysRevLett.115.087002}

{\bf 3. 2: Non-magnetic Weyl Semimetal: TaAs family}
\hspace*{2pt}


In addition to the two magnetic WSMs, namely pyrochlore iridate 
$Rn_2$Ir$_2$O$_7$~\cite{WanXG_WeylTI_2011} and spinel HgCr$_2$Se$_4$~\cite{XuGang_HgCrSe_2011_PRL}
introduced above, other proposals involve a fine-tuned multilayer structure composed of normal insulator and magnetically 
doped TIs~\cite{Burkov_Weyl_2011} or just magnetic ion doped TIs.~\cite{PhysRevB.89.081106} 
All of them involve magnetic compounds and the spin degeneracy of the bands is removed 
by breaking time reversal symmetry. However, these proposals 
are still waiting for experimental confirmation. 

As mentioned above, WSM can also be generated by breaking the spatial inversion symmetry only but keeping
time-reversal symmetry. This is the nonmagnetic and noncentrosymmetric WSM. Compared with the magnetic
WSM, nonmagnetic WSM has the following advantages. 1) it can be more easily studied by 
ARPES measurement since the alignment of magnetic 
domains is not required; 2) without the spin exchange field, the unique structure of Berry curvature 
related with magnetic monopoles leads to very unusual transport properties under external magnetic field. Such 
properties are not mixed by the self-magnetization of the magnetic sample. 

There have been several proposals for WSM in system without inversion symmetry, such as a 
superlattice composed of alternatively stacking normal and topological insulators,~\cite{Burkov_Weyl_2011, Burkov_Weyl_semimetal_2012PRB} 
elemental Tellurium or Selenium crystal under proper pressure,~\cite{SeTe} the fine-tuned solid 
solution of $A$Bi$_{1-x}$Sb$_x$Te$_3$ ($A$=La and Lu)~\cite{WS_Vanderbilt_2014} around
the topological transition points.~\cite{murakami_phase_2007, Murakami2011748} 
There is none of the above has been realized experimentally due to the following 
possible problems: 1) very fine control, such as quantum well thickness and doping
concentration, is required in sample preparation; 2) applying pressure makes the other measurement very difficult.

In the end of 2014,  our collaborators and us have posted the theoretical prediction that TaAs,TaP, NbAs and NbP 
single crystals are natural WSM and possess 12 pairs of Weyl points.~\cite{TaAs_Weng}
Compared with the other proposals, this family of materials are completely stoichiometric without any additional
doping, external strain or pressure. High quality sample can be easily obtained and experimental verification
becomes quite easier.

Unlike in the case of pyrochlore iridates and HgCr$_2$Se$_4$, inversion symmetry is kept and the 
appearance of Weyl points can be immediately inferred from the product of the parities at all the time 
reversal invariant momenta (TRIM).~\cite{FuLiang_3dTI_2007PRL, turner_quantized_2012, inversion2} 
For the TaAs family, parity is no longer a good quantum number. It is found that the appearance of 
Weyl points can be inferred by analyzing the mirror Chern numbers (MCN)~\cite{Hsieh_EXP_TCI_2012, WengHM_YbB6_2014PRL} 
and $Z_2$ indices~\cite{Kane_Z2_2005, FuLiang_Z2Pump_2006,FuLiang_3dTI_2007PRL,FuLiang_TI_with_inversionSymmetry_2007} for the four mirror and time reversal invariant planes in 
the BZ. Similar as many other topological materials, the WSM phase in this family is also induced by 
band inversion, which, in the absence of spin-orbit coupling (SOC), leads to nodal rings in 
the mirror plane. Once SOC is turned on, each nodal ring will be gapped with the exception of 
three pairs of Weyl points, which lead to fascinating physical properties including the complicated 
Fermi arc structures on surface.

\begin{figure}[tbp]
\includegraphics[scale=0.4]{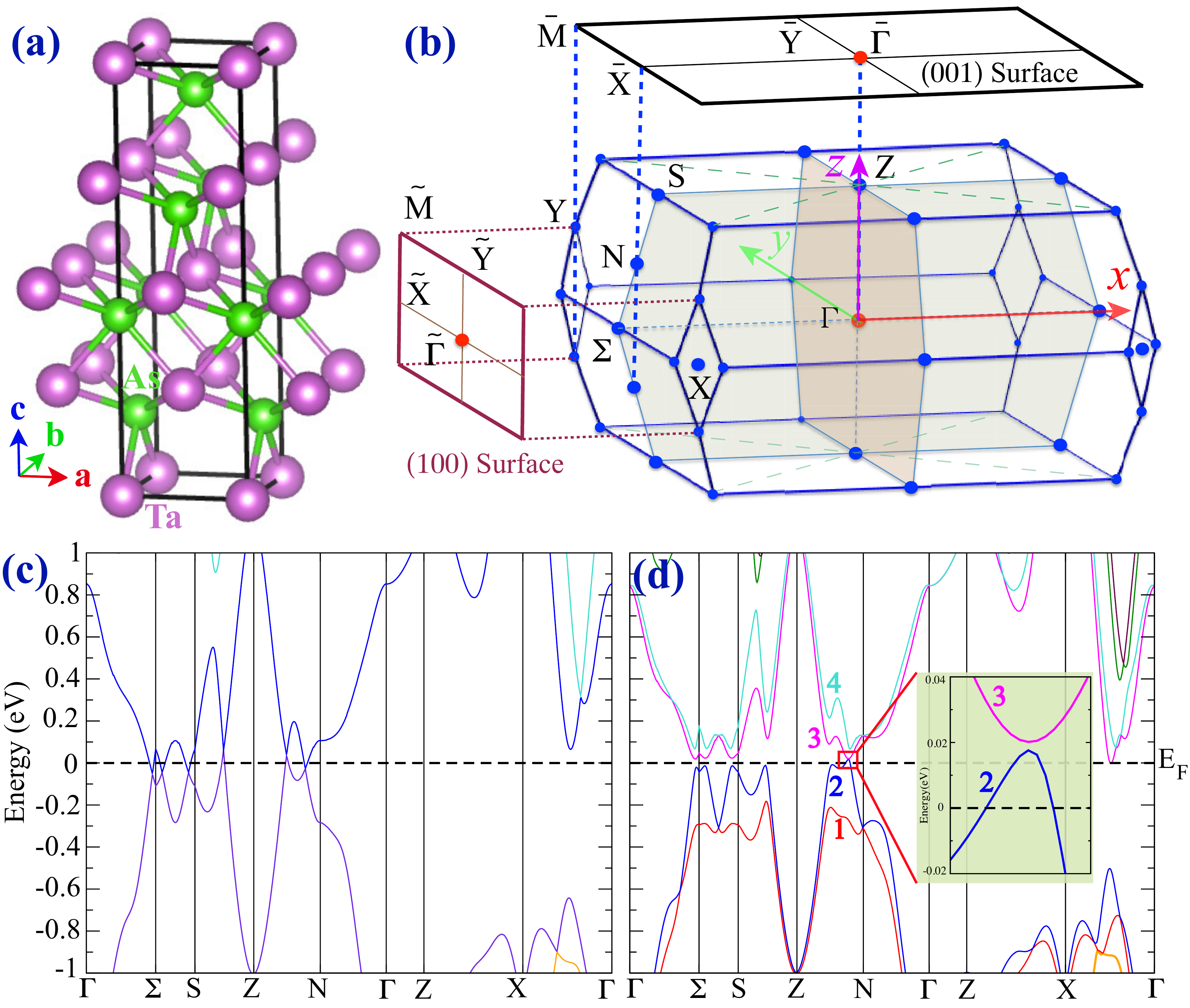}
\caption{{\bf Crystal structure and Brillouin zone (BZ)}. (a) The crystal symmetry of TaAs; 
(b) The bulk BZ and the projected surface BZ for both (001) and (100) surfaces; 
(c) The band structure of TaAs calculated by GGA without including the spin-orbit coupling; 
(d)The band structure of TaAs calculated by GGA with the spin-orbit coupling.}
\label{TaAs_crystal_structure}
\end{figure}

The experimental crystal structure of TaAs~\cite{expCrystStru} is shown in Fig. ~\ref{TaAs_crystal_structure}. 
It has body-centered-tetragonal structure with nonsymmorphic space group $I4_{1}md$ (No.~109) lacking 
inversion symmetry. The measured lattice constants are $a$=$b$=3.4348~\AA~and $c$=11.641 \AA. 
Both Ta and As are at $4a$ Wyckoff position (0, 0, $u$) with $u$=0 and 0.417 for Ta and As, respectively.
To calculate the topological invariant such as MCN and surface states of TaAs, we have generated atomic-like 
Wannier functions for Ta $5d$ and As $4p$ orbitals using the scheme described in Ref.~\onlinecite{myMLWF} 


\begin{figure}[tbp]
\includegraphics[width=0.75\textwidth]{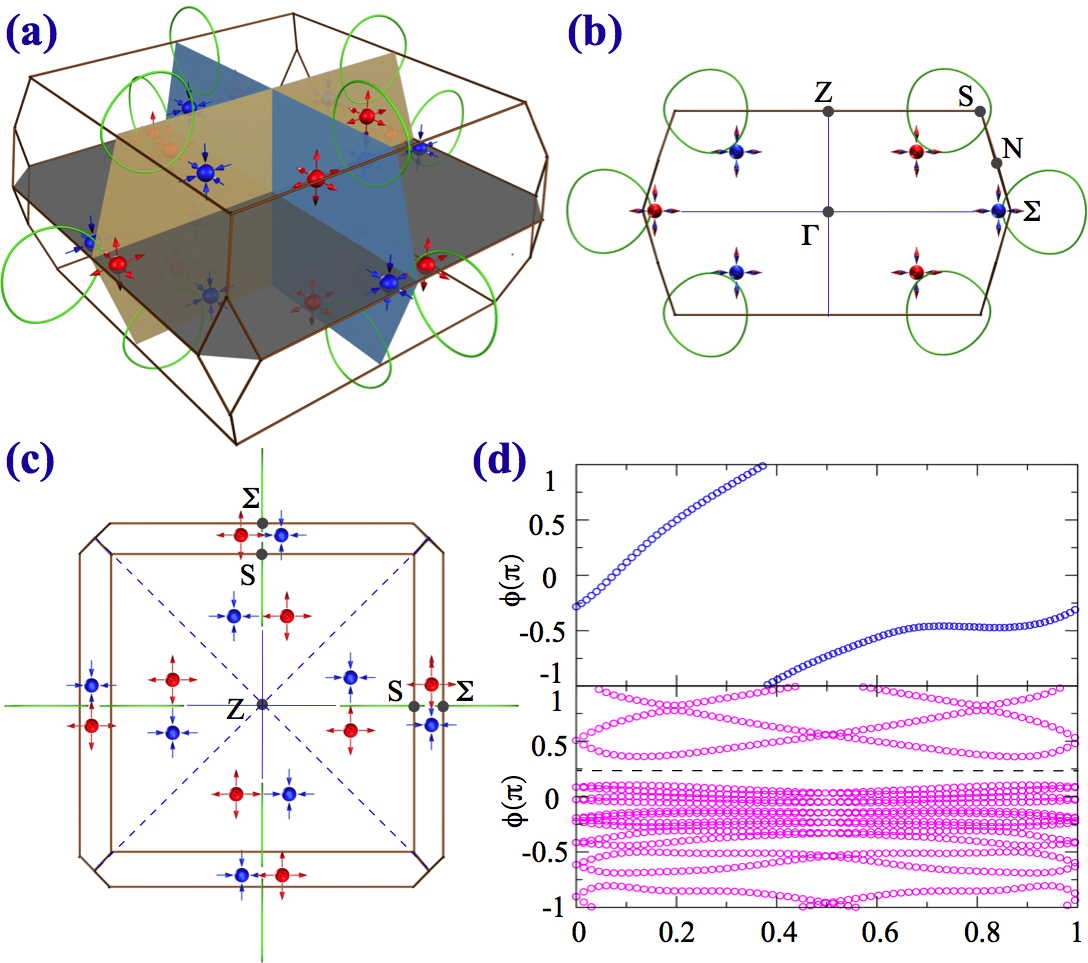}
\caption{{\bf Nodal rings and Weyl points distribution, as well as $Z_2$ and MCN for mirror planes}. (a) 3D view of 
the nodal rings (in the absence of SOC) and Weyl points (with SOC) in the BZ; (b) Side view from [100] and (c) top 
view from [001] directions for the nodal rings and Weyl points. Once the SOC is turned on, the nodal rings are 
gapped and give rise to Weyl points off the mirror planes; (d) top panel: Flow chart of the average position of the 
Wannier centers obtained by Wilson loop calculation for bands with mirror eigenvalue $i$ in the mirror 
plane ZN$\Gamma$; bottom panel: The flow chart of the Wannier centers obtained by Wilson loop calculation 
for bands in the glide mirror plane ZX$\Gamma$. There is no crossing along the reference line (the dashed line) 
indicating the $Z_2$ index is even.
}
\label{TaAs_band_structure}
\end{figure}

The band structure of TaAs within GGA without including SOC is shown in Fig.~\ref{TaAs_crystal_structure}(c). 
There is clear band inversion and multiple band crossing features near the Fermi level along ZN, ZS and $\Sigma$S 
paths. In Fig.\ref{TaAs_crystal_structure}(b), the shadowed planes are two mirror planes, namely $M_{x}$, $M_{y}$, while
the dashed lines indicate two glide mirror planes, namely $M_{xy}$, $M_{-xy}$.
The plane spanned by Z, N and $\Gamma$ points is invariant under mirror $M_y$ and the energy bands within the plane can 
be labeled by mirror eigenvalues $\pm 1$. Further analysis shows that the two bands that cross the Z to N 
line belong to opposite mirror eigenvalues and hence the crossing between them is protected by mirror symmetry. Similar 
band crossings can also be found along other high symmetrical lines in the ZN$\Gamma$ plane, such as ZS and NS. 
Therefore, these band crossing points form  a ``nodal ring" in the ZN$\Gamma$ plane as shown in Fig.\ref{TaAs_band_structure}(b). 
Unlike for the situation in ZN$\Gamma$ plane, in the two glide mirror planes ($M_{xy}$ and $M_{-xy}$), the band structure 
is fully gaped with the minimum gap around 0.5 eV.


The bands near the Fermi energy are mainly formed by Ta $5d$ orbitals, which have quite strong SOC. 
SOC brings dramatic changes in the band structure near Fermi level, as plotted in Fig.~\ref{TaAs_crystal_structure}(d). 
It seems that the previous band crossings on the nodal rings in ZN$\Gamma$ plane are all gaped. Detailed 
symmetry analysis reveals that the bands ``2" and ``3" in Fig.\ref{TaAs_crystal_structure}(d) belong 
to opposite mirror eigenvalues, indicating the almost touching point along the ZN line is completely 
accidental. In fact there is a small gap of roughly  3 meV between bands ``2" and ``3" as illustrated by 
the inset of Fig.\ref{TaAs_crystal_structure}(d). Therefore, the ZN$\Gamma$ plane becomes fully 
gapped with SOC included.

There are three critical questions to be addressed to convincingly identify TaAs as a WSM. 1) Do Weyl nodes really exist? 
2) How many of them? 3) Where are they? Since Weyl nodes are at general $k$-points with accidental double degenerate 
eigen energy, numerical method always has error due to discrete sampling of 3D momentum space. No matter how dense 
the sampling grid is used, finite numerical error exists and brings difficulty in identifying the existence of massless Dirac 
cone like band structure. Some efficient way mathematically rigorous is necessary to make final conclusion. 

Since TaAs has no inversion center, the widely used and most simplified parity configuration method~\cite{FuLiang_3dTI_2007PRL, FuLiang_TI_with_inversionSymmetry_2007, turner_quantized_2012, inversion2} 
can not be applied to identify the existence of WSM. Another strategy is thus developed in Ref.~\onlinecite{TaAs_Weng}. 
As previously mentioned, the space group of the material provides two mirror planes ($M_x$ and $M_y$), where 
MCN can be defined. If a full gap exists for the entire BZ, the MCN would directly reveal whether this system 
is a topological crystalline insulator~\cite{FuLiang_Topological_Crystalline_2011PRL} or not. Interestingly, 
if the system is not fully gaped MCN is still useful to find out whether the material hosts Weyl points in the 
BZ or not. Besides the two mirror planes, there are two additional glide mirror plane ($M_{xy}$ and $M_{-xy}$).
MCN is not well defined for the glide mirror planes, but $Z_2$ index is still well defined here
since these two planes are time reversal invariant and electronic states inside of them are insulator. The 
Wilson loop method~\cite{YuRui_Z2_2011PRB, MRS_review} is applied to calculate 
the MCNs for the two mirror planes and $Z_2$ indices for the two glide mirror planes. 
To determine the MCNs for the mirror plane $M_x$, Wilson loops along $k_x$ direction with fixed $k_z$ are defined. 
All the occupied bands at $k$-points in this plane can be classified into two groups according to their eigenvalues under 
mirror operation, $i$ or $-i$. For the bands having eigenvalue $i$, the evolution of Berry phases along periodic $k_z$ 
direction can be obtained and the MCN is simply its winding number. In Fig.~\ref{TaAs_band_structure}(d), it is clear that
MCN is one for ZN$\Gamma$ plane ($M_y$) and $Z_2$ index is even for ZX$\Gamma$ plane($M_{xy}$). Thus, for the 
$(001)$ surface, which is invariant under the $M_y$ or $M_x$ mirror, there will appear non-trivial helical edge modes
due to non-zero MCN in  ZN$\Gamma$ plane, which generates a single pair of
FS cuts along the projective line of the ZN$\Gamma$ plane, i.e. the $x$- or $y$-axis in Fig.~\ref{TaAs_band_structure}(c).
Whether these Fermi cuts will eventually form a single closed Fermi circle or not 
depends on the $Z_2$ index for the two glide mirror plane, which are projected to the dashed blue lines in
Fig.~\ref{TaAs_band_structure}(c). Since the $Z_2$ indices for the glide mirror planes are trivial, there are 
no protected helical edge modes along the projective lines of the 
glide mirror planes, the dashed blue lines in Fig.~\ref{TaAs_band_structure}(c), and the Fermi level cuts 
along the $x$- or $y$-axis in Fig.~\ref{TaAs_band_structure}(c) must end somewhere between the $x$- or $y-$ axis 
and the diagonal lines. That means they must be Fermi arcs, ending at the surface projection of Weyl points in the 
bulk band structure of TaAs.

\begin{figure}[tbp]
\includegraphics[width=0.40\textwidth]{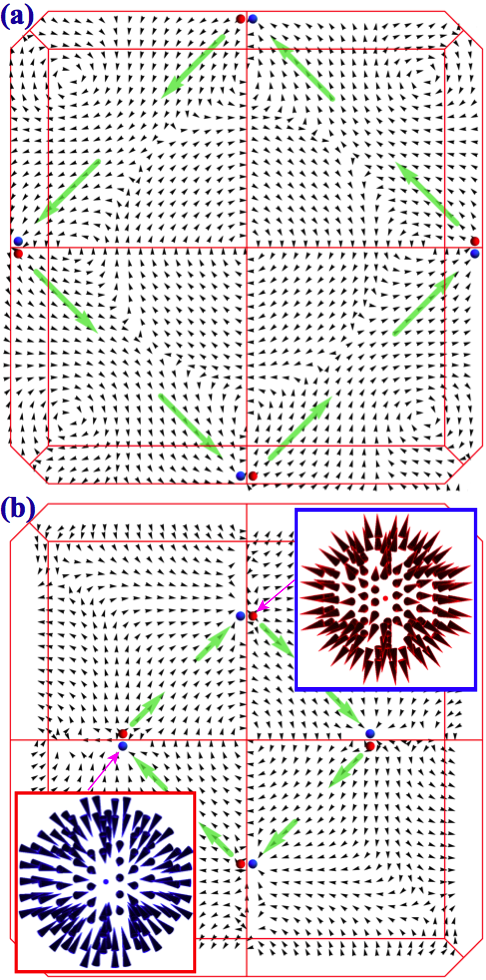}
\caption{{\bf Berry curvature from pairs of Weyl points}. (a) The distribution of the Berry curvature for the $k_z=0$ plane, where the blue and red dots denote the Weyl points with chirality of +1 and -1, respectively; (b) same with (a) but for  $k_z= 0.592\pi$ plane. The insets show the 3D view of hedgehog like Berry curvature near the two selected Weyl points.}
\label{TaAs_Berry_Curvature}
\end{figure}

The existence of Weyl points is confirmed. The left questions are the total number and the exact positions of them. 
As mentioned above, this is a hard task since the Weyl points are located at generic $k$ points without any 
symmetry. For this purpose, we calculate the integral of the Berry curvature on a closed surface in $k$-space, which 
equals to the net chirality of the Weyl nodes enclosed by the given compact surface. Due to the four fold rotational symmetry 
and mirror planes that characterize TaAs, one only needs to search for the Weyl points within the reduced BZ, i.e., 
one eighth of the whole BZ. The total chirality or monopole charge enclosed in the whole reduced BZ is found to be
one. This guarantees the existence of, and odd number of Weyl points in one eighth of BZ.
 To determine the exact location of each Weyl point, the reduced BZ is divided into very dense $k$-grid and the 
Berry curvature or the ``magnetic field in momentum space"~\cite{fang_anomalous_2003, WangXJ_ab_AHE_Wannier_2006, myMLWF} 
is calculated as shown in Fig.~\ref{TaAs_Berry_Curvature}. From this, one can easily identify the precise positions of the Weyl points 
by searching for the ``source" and ``drain" points of the ``magnetic field". Around the Weyl points, the hedgehog like Berry curvature
is clearly shown in Fig.~\ref{TaAs_Berry_Curvature}. 

Therefore, there are totally twelve pairs of Weyl points in TaAs as illustrated in Fig.~\ref{TaAs_band_structure}(a). All
of them are in the vicinity of the nodal rings on two of the mirror invariant planes in the SOC-free case. There are two
types of Weyl nodes. W2 are those located off the $k_z=0$ plane, totally eight pairs. W2 is those located exactly in 
the $k_z=0$ plane, totally four pairs.  All W1 Weyl nodes are related by mirror symmetry, four-fold rotational symmetry,
and time reversal symmetry, so does it for W2. However, W1 and W2 are not related by any symmetry and they
can have different velocity matrix and located at different energy. For example,  in TaAs, W2 nodes are about 2 meV 
above the Fermi energy and form eight tiny hole pockets, while the W1 nodes are about 21 meV below the Fermi level 
to form sixteen electron pockets. These values are obtained by dense $k$-grid band structure calculation based on
tight-binding model from MLWFs of Ta $d$ and Te $p$ orbitals. SOC is included as onsite terms with
0.3 eV for Ta $d$ and 0.2 eV for Te $p$ as fitted from first-principles calculation with SOC included.
 The band structures for the other  three materials TaP, NbAs and NbP are very similar. The precise positions of the 
 Weyl points for all these materials are summarized in Table.~\ref{weylpoint}.

 
 \begin{table}

\caption{
 \label{weylpoint}
The two nonequivalent Weyl points in the $xyz$ coordinates shown in Fig.~\ref{TaAs_crystal_structure}(b). 
The position is given in unit of the length of $\Gamma$-$\Sigma$ for $x$ and $y$ and of the length of $\Gamma$-Z for $z$.}
\begin{tabular*}{0.45\textwidth}{@{\extracolsep{\fill}}c|c|c}
 	\hline\hline
    & W1 & W2  \\
    \hline
TaAs & (0.520, 0.037, 0.592) & (0.949, 0.014, 0.0)  \\
 	\hline
TaP & (0.499, 0.045, 0.578) & (0.955, 0.025, 0.0) \\
 	\hline
NbAs & (0.510, 0.011, 0.593) & (0.894, 0.007, 0.0) \\
 	\hline
NbP  & (0.494, 0.010, 0.579) & (0.914, 0.006, 0.0) \\
 	\hline\hline
\end{tabular*}
\end{table}

One of the hallmark of WSM is the Fermi arcs on the surface.~\cite{WanXG_WeylTI_2011, XuGang_HgCrSe_2011_PRL, Transport_Weyl_XLQi_2013}
In TaAs family, there are multiple Weyl points. When they are projected to one surface, some of them might be 
on top of each other. It is easy to see that the total number of surface modes at Fermi level crossing 
a closed circle in surface BZ must equal to the sum of the ``monopole charge" of the Weyl points inside the 
3D cylinder that projects to the given closed circle. Another important fact determining the behavior 
of the surface states is the MCN introduced in the above discussion, which limits the number of FSs 
cutting certain projection lines of the mirror plane as long as this mirror symmetry is still preserved on 
the surface.

\begin{figure}[tbp]
\includegraphics[width=0.8\textwidth]{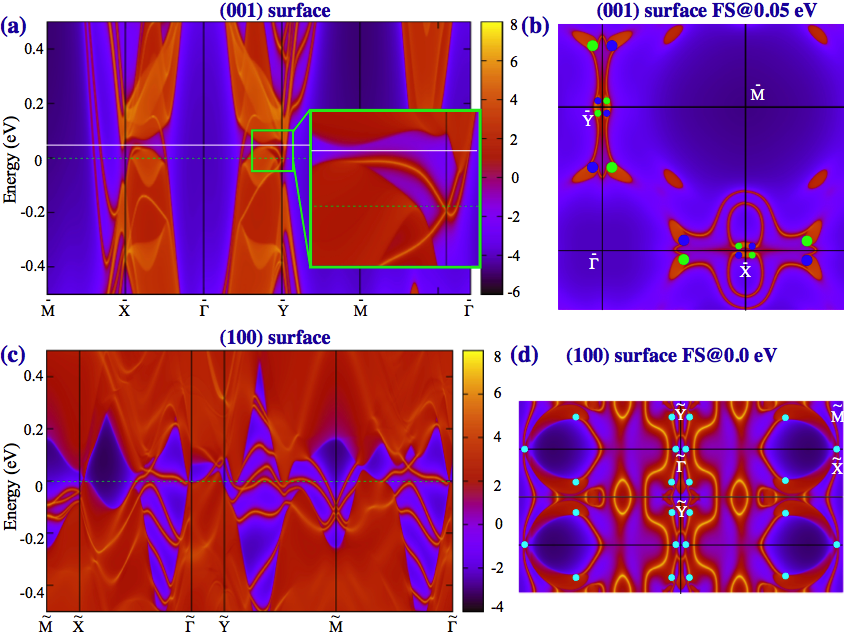}
\caption{{\bf Fermi arcs in the surface states}. (a) The Surface states for (001) surface; (b) the corresponding fermi surfaces on (001) surfaces; 
(c) The Surface states for (100) surface. The dots illustrate the 
projective points of the bulk Weyl points on the surfaces, where the color represents the 
chirality of the Weyl points (blue for positive and green for negative), 
small dot represents the single projected Wely point, and
large dot represents two Weyl points with same chirality projecting
on top of each other.
(d) the corresponding fermi surfaces on (100) surfaces. All the dots here are the projective points for a pair of Weyl points with opposite chirality.  }
\label{TaAs_surf_state}
\end{figure}

By using the Green's function method~\cite{MRS_review}, the surface states for both (001) and (100) 
surfaces have been calculated as shown in Fig.~\ref{TaAs_surf_state} together with its Fermi surface. 
On the (001) surface, the crystal symmetry is reduced to $C_{2v}$ leading to different behavior for 
the surface bands around $\bar X$ and $\bar Y$ points respectively. 
Along the $\bar\Gamma$-$\bar X$ or $\bar\Gamma$-$\bar Y$ lines, 
Fermi surface crosses it twice with opposite Fermi velocity. This satisfies the constraint 
from the MCN for $\Gamma$ZN plane. In addition to the MCN, the possible 
``connectivity pattern" of the Fermi arcs on the surface has to link different
projection points of the Weyl nodes with opposite chirality. 
However, for the (001) surface of TaAs, the connectivity pattern of the Fermi arcs 
that satisfy all the conditions discussed is not unique. Only the appearance of ``Fermi arcs" 
on the (001) surface is guaranteed since all the projective points on the (001) surface are 
generated either by a single Weyl point or by two Weyl points with the same chirality. 
The Fermi arc connectivity pattern for (001) surface shown in Fig.~\ref{TaAs_surf_state}(b) is
obtained from our $ab$-$initio$ calculation on a non-relaxed surface described by the TB model
based on MLWFs of Ta $d$ and Te $p$ orbitals. Changes of surface potentials or the simple 
relaxation of the surface charge density might lead to modification of the Fermi arc 
connectivity pattern and result in topological Fermi arc phase transitions on the surface. 
The extremely long Fermi arcs which cross the zone boundary along the $\bar X$ to $\bar M$ line
is quite interesting. Comparing with other proposed WSM materials, the Fermi arcs 
in TaAs families are much longer, which greatly facilitates their detection in experiments. 

On the other hand, the surface states on (100) surface of TaAs are much more complicated 
as shown in Fig.~\ref{TaAs_surf_state}(c) and (d). The biggest difference between them is 
that all the projected Weyl points on the (100) surface are formed by a pair of Weyl points 
with opposite chirality, which does not guarantee (but does not disallow) the existence of 
the Fermi arcs. The only constraint for the (100) surface states is the nonzero MCN of 
the $\Gamma ZN$ plane, which generates a pair of chiral modes along the
$\bar\Gamma \bar Y$ line, the projection of the mirror plane, as illustrated 
in Fig.\ref{TaAs_surf_state}(d).

After this theoretical perdition, our collaborators and us have reported several experimental works to confirm it. 
Totally, there are four hallmarks of WSM in TaAs family have been observed: 1) Fermi arcs on its (001) surface;~\cite{PhysRevX.5.031013} 2) Weyl nodes in 
its bulk state;~\cite{TaAs_NatPhys2015} 3) negative magnetoresistivity due to ``chiral anomaly";~\cite{PhysRevX.5.031023} 
4) Spin texture of the Fermi arcs on (001) surface.~\cite{PhysRevLett.115.217601} The other 
experimental works appeared include Ref.~\onlinecite{Yang2015, Liu2015aj, Shekhar2015,Jia2015PRB, Xu2015a,Xu2015b,Xu2015c,Xu2015d, 2015arXiv150703983X,FengDL2015,HuangSM_Weyl}.
The topological superconductivity~\cite{Hosur2014pn, Jian2015, Li2015yi,PhysRevB.92.035153} is also expected for either DSM or WSM, 
and several experimental efforts~\cite{WangJ2015super, LiSY2015super,ZHANGJun:97102} have been reported.

{\large\bf 4. Node-Line Semimetal}
\hspace*{2pt}

In the former two sections, we have discussed the TSMs with isolated Dirac cone like nodes at or close to the Fermi level. 
However, there is another interesting electronic states having the nodes forming a closed ring 
or periodically continuous line in momentum space. Such kind of semimetal state now is known as topological nodal 
semimetal~\cite{Burkov_Topological_nodal_semimetals_2011PRB} or topological Node-Line 
semimetal (NLSM).~\cite{allcarbon_nodeLine2014} In 2011, Burkov {\it et al.} proposed that 
a fine tuned superlattice of normal insulator and TI with broken TRS might realize 
NLSM.~\cite{Burkov_Topological_nodal_semimetals_2011PRB}  This
is basically a model proposal rather than a practically realizable material candidate. In this section, 
we will introduce some realistic 3D compounds proposed to host NLSM. The
underlying mechanism is also based on band inversion, which causes Node-Line when both TRS
and IS are conserved and SOC is neglected for light element compounds.

{\bf 4. 1: All Carbon Mackay-Terrones Crystal}
\hspace*{2pt}

\begin{figure}
\includegraphics[scale=0.6]{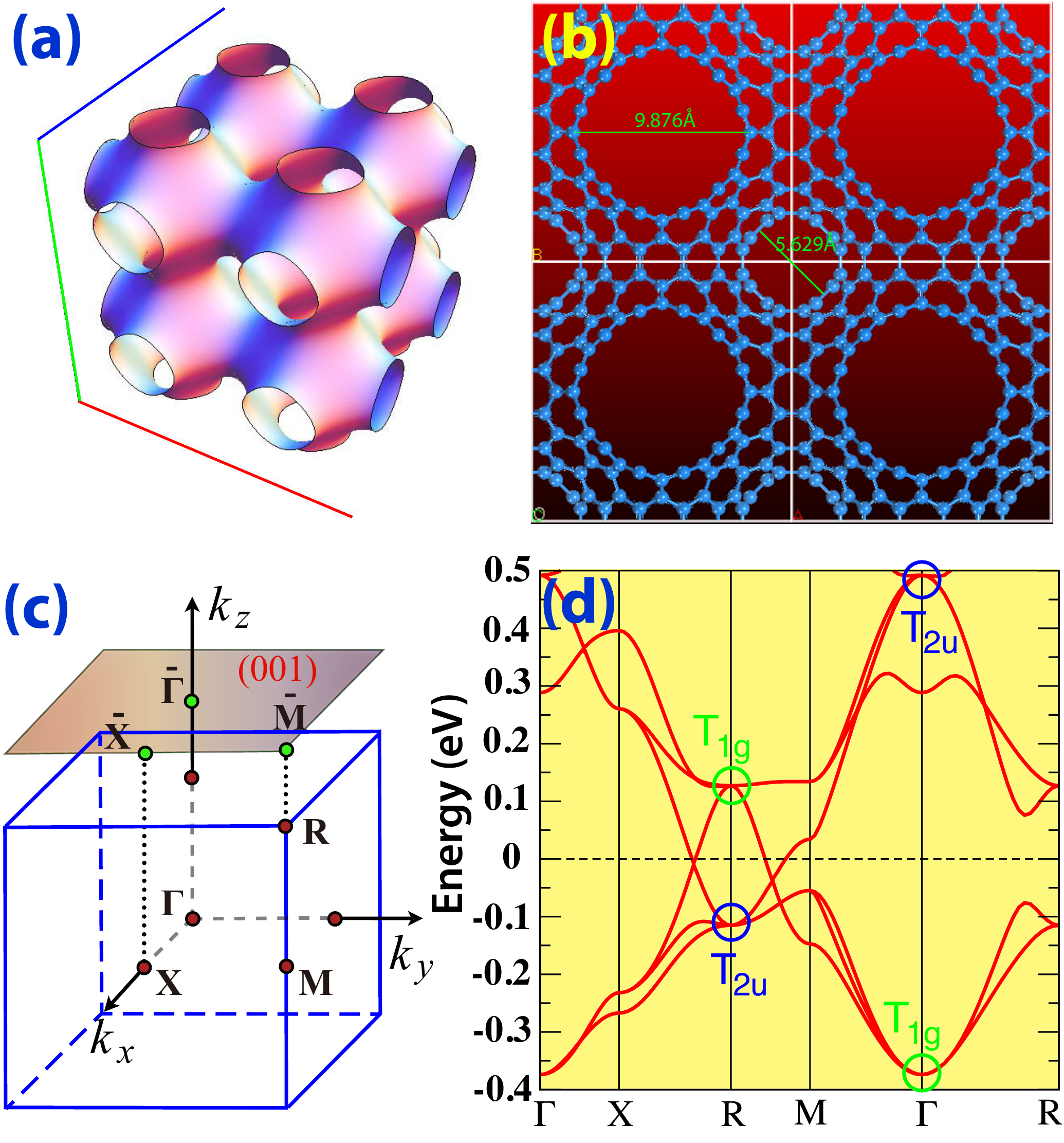}
\caption{(Color online) (a) The Schwarz minimal P-surface in 2$\times$2$\times$2 supercell. 
(b) The top view of 6-1-1-p MTC in 2$\times$2 supercell. (c) Bulk and (001)-surface Brillouin 
zone, as well as the high symmetrical k-points. (d) Band structure from the first-principles 
calculation. The two triply degenerated eigenstates at $\Gamma$ and R with T$_{1g}$ 
and T$_{2u}$ symmetrical representation are marked. The band 
inversion between them can be easily seen.}
\label{fig:MTCstructure}
\end{figure}

In 1992, A. L. Mackay and H. Terrones proposed that~\cite{mackay} graphene can 
be extended to form 3D networks by placing graphitic tiles 
consisting of 4- to 8-membered rings onto the Schwarz minimal surface. 
Such kind of 3D all carbon allotrope is called as
Mackay-Terrones crystal (MTC). Schwarz minimal surface is a 3D periodic 
minimal surface with its mean curvature $H=(k_1+k_2)/2$ being zero 
and Gaussian curvature ($K=k_1k_2$) being negative everywhere on it. 
The above $k_1$ and $k_2$ are the principal curvatures. There are various 
Schwarz minimal surfaces, such as primitive (P), diamond (D) and gyroid (G) 
surface. One type of MTC based on P-surface is shown in Fig.~\ref{fig:MTCstructure}. 
Different from C$_{60}$-like fullerene, which has positive Gaussian 
curvature, MTC has negative Gaussian curvature and is periodically 
connected. It is shown such all-carbon MTC can host non-trivial electronic states, including 
topological node-lines and 3D Dirac points, which are distinct from its 2D counter 
material graphene. Similar node-lines have also been proposed in optimally tuned photonic 
crystal composed of gyroid,~\cite{lufu} the Schwarz minimal G-surface.



\begin{figure}
\includegraphics[scale=0.7]{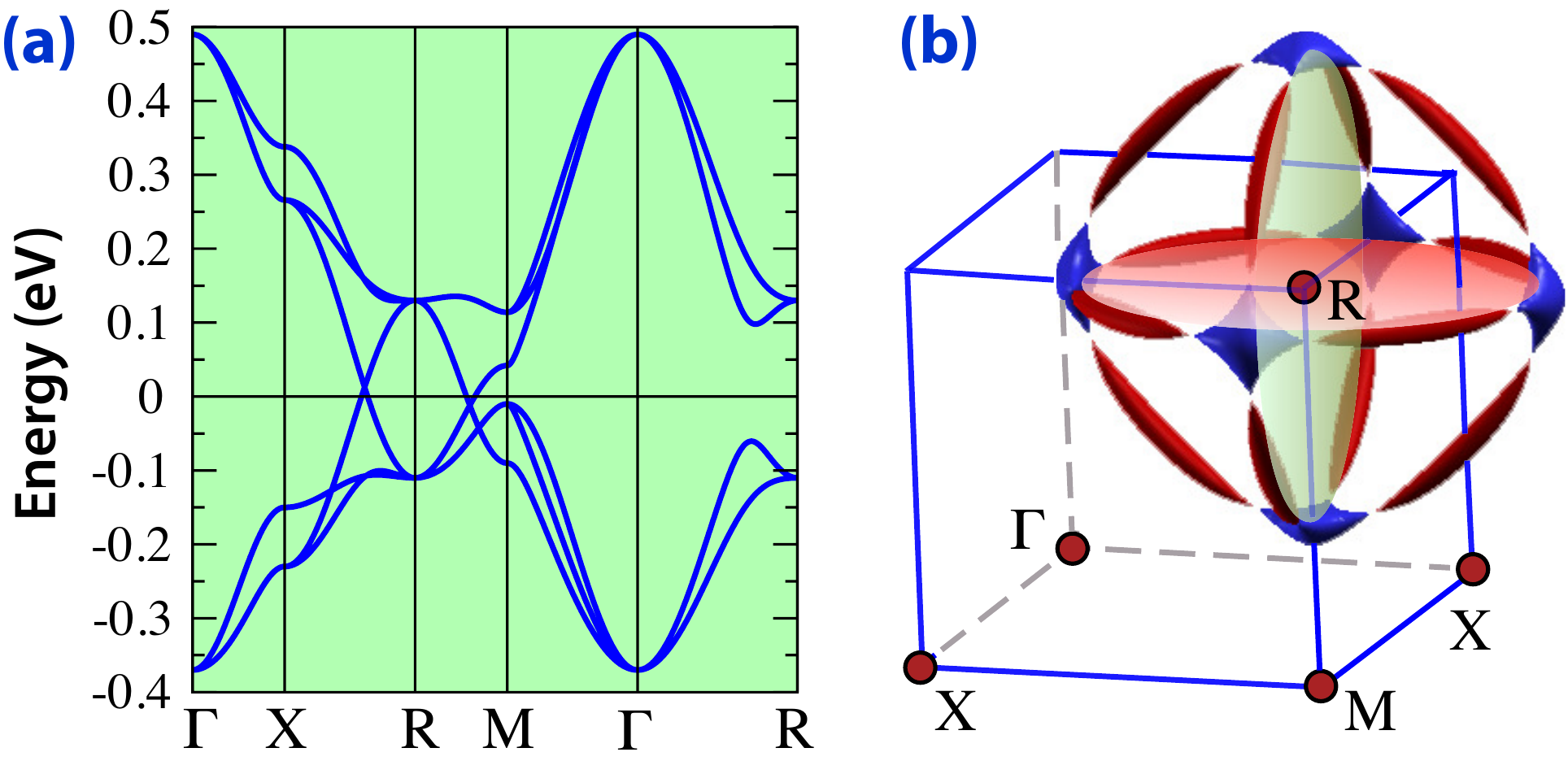}
\caption{(Color online) (a) Band structure from effective tight-binding model calculation, 
which reproduces all the features of Fig.~\ref{fig:MTCstructure}(d). (b) Fermi surface consists of three lotus 
root like rings. These rings are centering R point and parallel to the $k_x$=$\frac{\pi}{a}$, 
$k_y$=$\frac{\pi}{a}$ and $k_z$=$\frac{\pi}{a}$ plane, respectively. They are formed by
the electron pockets (blue) and hole pockets (red) connected by nodal points at Fermi energy. 
}
\label{fig:Modelband}
\end{figure}

Fig.~\ref{fig:MTCstructure} shows a stable structure with simple cubic lattice in
$Pm\bar{3}m$ space group. It contains 176 atoms per unit cell and is obtained
by Tagami {\it et al.} in Ref.~\onlinecite{ours} and labeled as 6-1-1-p. 
The electronic band structure of this crystal, calculated based on the local 
density approximation (LDA), is shown in Fig.~\ref{fig:MTCstructure}(d). 
It indicates this crystal is a semimetal with band crossings around the 
Fermi level, similar to the massless Dirac cone in graphene, but they are essentially 
different. The occupied and unoccupied low energy bands are triply-degenerate at $\Gamma$, and
have T$_{1g}$ (even parity) and T$_{2u}$ (odd parity) symmetry, respectively. 
Moving away from $\Gamma$ point, their degeneracy is lifted but at R point, 
their degeneracy is recovered again. However, the odd and even eigenstates 
exchanges their energy ordering, which is a typical band inversion picture.
Thus, the band inversion leads to the band crossings along both $X-R$ and $R-M$
paths as seen from Fig.~\ref{fig:MTCstructure}(d). Including SOC will open a 
band gap on the nodal points and it becomes a 3D strong TI
with $Z_2$ index (1;111)~\cite{FuLiang_3dTI_2007PRL, FuLiang_TI_with_inversionSymmetry_2007} if the lower (upper) half of the anti-crossing bands
are thought as occupied (unoccupied). However, the computed SOC splitting 
is very small, around 0.13 meV or equivalent to 1.5 K temperature, which 
can be neglected when temperature is higher than 1.5 K.

Without SOC, the most interesting thing found in MTC is that the band crossing
points lead to nodal lines rather than nodal points. In fact, the band crossings 
form closed loops in the 3D momentum space, and generate three circular-like
node-lines around the $R$ point, as shown in Fig.~\ref{fig:Modelband}. 

Detailed analysis indicates that these node-lines are protected by two factors.
One is the coexistence of TRS and IS and the other is that the SOC is negligible. 
The cubic symmetry of MTC leads to three in-plane nodal rings. The node-lines
are not necessarily to be flat in energy, i. e., the crossing points might be at 
different energy level at different $k$, which leads to compensated 
electron and hole pockets along the node-lines. 
 
 Different from other proposals for the topological node-lines, such as that 
 in Ref.~\onlinecite{Burkov_Topological_nodal_semimetals_2011PRB}, 
 the appearance of node-lines in MTC is very stable and does not require 
 fine tuning of any parameter. This mechanism to generate topological node-lines 
 in 3D materials only requires TRS, IS and negligible SOC, 
 which can be easily applied to a large class of materials consisting of mainly 
 the light elements.

\begin{figure}
\includegraphics[scale=0.5]{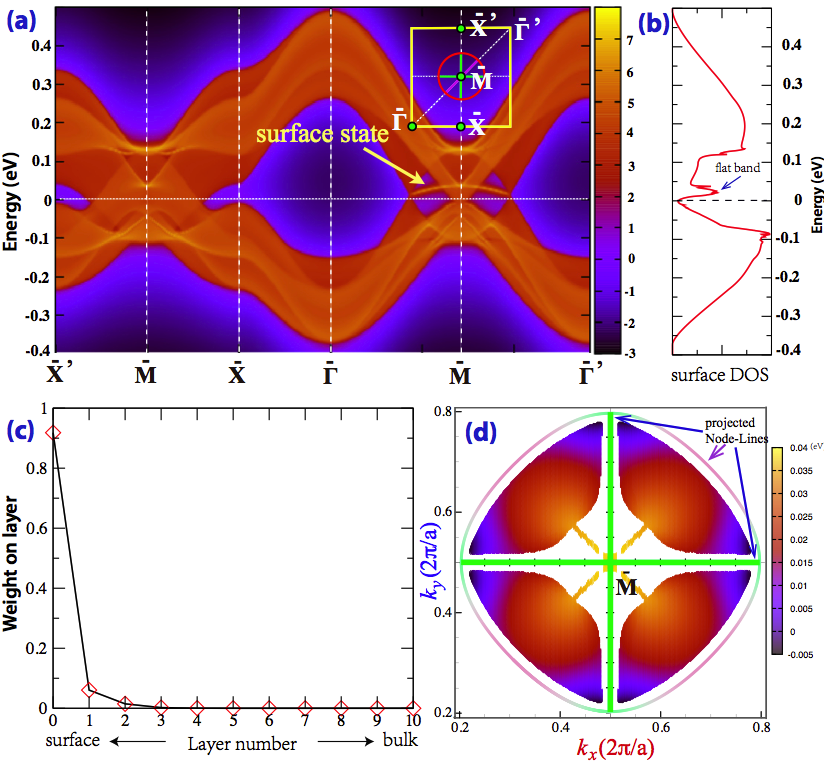}
\caption{(Color online) The (001)-surface state. (a) The nearly flat surface band is nestled between two solid Dirac cones, 
which are the projection of one of the node-line circles as indicated in the inset (red circle). The other two node-line rings 
are projected as two orthogonal diameters (green line). (b) The surface density of state. (c) The wave function of surface state pointed
by the arrow decay rapidly into bulk. (d) The eigen energy distribution of surface flat band nestled inside of projected node-line circle, 
which looks like a vibration model of ``drumhead". The mixing of surface and bulk state leads to discontinuous in this plot.
}
\label{fig:MTC_surf}
\end{figure}

This topologically stable node-line semimetal state can have nontrivial 
surface states.~\cite{Ryu_2002PRL, JETP932011, JETP942011, 2011arXiv1111.4627V, Burkov_Topological_nodal_semimetals_2011PRB} 
For the (001)-surface, the three node-line rings are projected to 
be a ring and two orthogonal diameter segments inside of it as shown in Fig.~\ref{fig:MTC_surf}(a). 
The (001)-surface state is calculated based on a six-band TB model using both Green's 
function method and slab-model method.~\cite{MRS_review}
There is a nearly flat surface band nestled inside of the projected node-line ring with 
its band width being about 40 meV due to the particle-hole asymmetry. The peak-like 
surface density of states contributed by this nearly flat band is clearly shown in Fig.~\ref{fig:MTC_surf}(b).
This is proposed to be an important route to high temperature surface superconductivity.~\cite{PhysRevB.83.220503, 2014arXiv1409.3944V} 
The weight of wave function of the surface flat band on each layer is shown in Fig.~\ref{fig:MTC_surf}(c). 
It penetrates just three layers into bulk with most of the weight on the surface layer. 
The surface localization of these flat bands is well resolved for those separated from 
bulk bands. The nestled flat surface states have small dispersion and 
their eigen energy distribution in surface BZ is shown in Fig.~\ref{fig:MTC_surf}(d), 
which looks like some vibrational mode of ``drumhead". Such ``drumhead"-like states 
are readily to be detected by angle-resolved photoelectron spectroscopy or scanning 
tunnel microscope. 

Recently, there have been many other efforts in studying the NLSM,~\cite{2015arXiv150502284C,LnX,Cu3NPd_Kane, Cu3NPd_Yu,CFang_NLSM_PRB,Ca3P2, PbTaSe2,TlTaSe2,BP_Zhao2015} including the theoretical proposal of Cu$_3$PdN to be introduced in the following.

{\bf 4. 2: Anti-Perovskite Cu$_3$PdN}
\hspace*{2pt}


In this section, cubic anti-perovskite Cu$_3$PdN is introduced.~\cite{Cu3NPd_Yu} 
It is another example system hosts Node-Line semimetal state protected by the
coexistence of TRS and IS in case with band inversion and negligible SOC. 
However, Pd is heavy element 
and quite strong SOC might open band gap along the nodal line as discussed
in Section 3. 2. Interestingly, in Cu$_3$PdN, the C$_4$ rotational
symmetry will protect the nodal points on the rotation axis and leads to DSM state.
This is totally different from the situation in TaAs, where SOC opens band
gap along the whole nodal ring but results in Weyl nodes off the nodal ring plane.
Thus, Cu$_3$PdN is a DSM with three pairs of Dirac nodes distinguished 
from Na$_3$Bi and Cd$_3$As$_2$ with only one pair as shown in Section 2. 1 and 2. 2, respectively.

\begin{figure}[t]
\begin{centering}
\includegraphics[width=0.8\columnwidth]{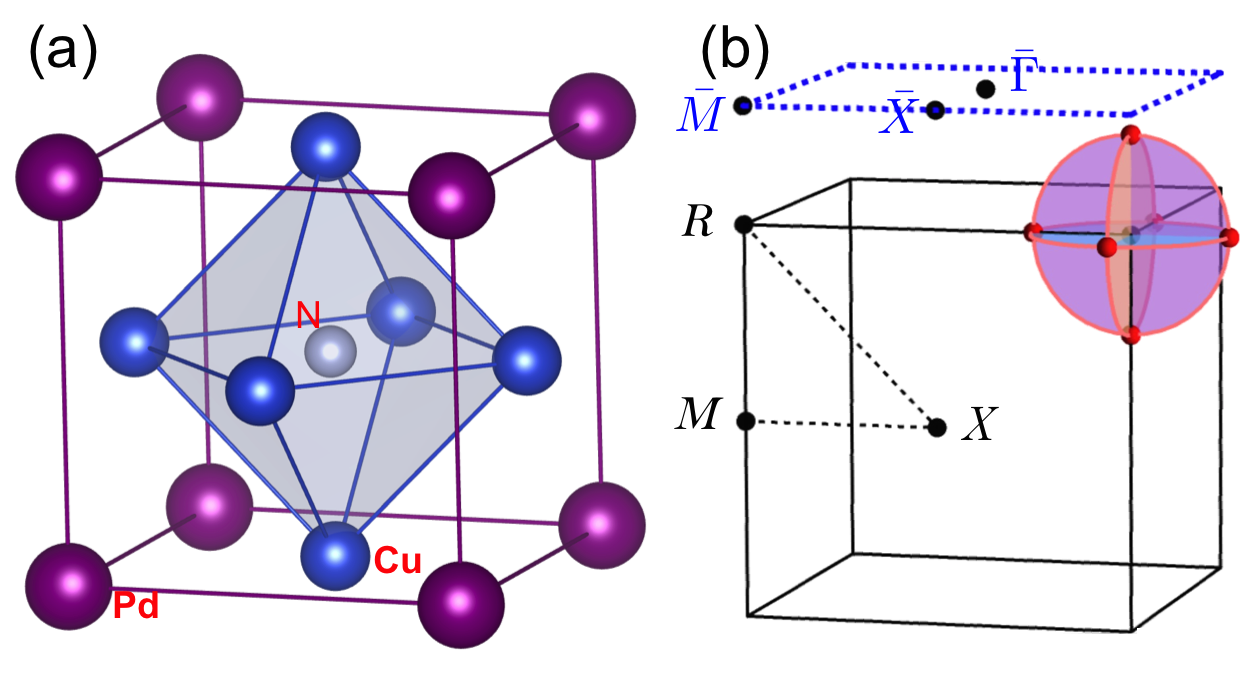}
\par\end{centering}
\protect\caption{\label{fig:crystal_BZ}(Color online)
(a) Crystal structure of anti-perovskite Cu$_3$PdN with Pm$\bar{3}$m (No.221) symmetry. 
Nitrogen atom is at the center of the cube and is surrounded by octahedral Cu atoms. 
Pd is located at the corner of the cube. (b) Bulk and projected (001) surface Brillouin zone. 
The three nodal line rings (orange color)  and three pairs of Dirac points (red points) 
without and with SOC included, respectively, are schematically shown.}
\end{figure}

\begin{figure}[t]
\begin{centering}
\includegraphics[width=0.8\columnwidth]{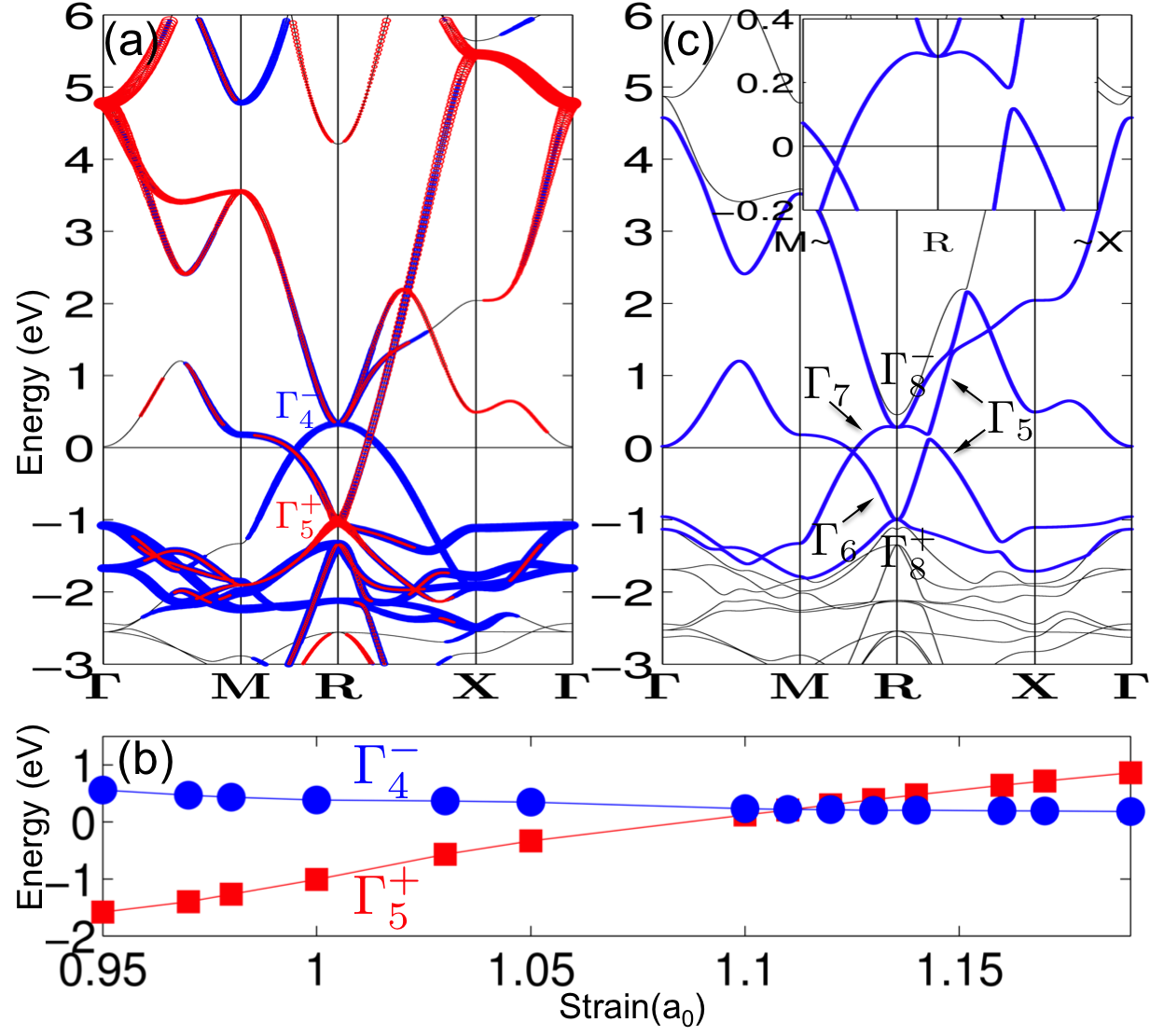}
\par\end{centering}
\protect\caption{\label{fig:bnd}(Color online)
(a) Electronic band structure without SOC, where the component of Pd-5$p$ (4$d$) 
orbitals is proportional to the width of the red (blue) curves. Band inversion 
between $p$ and $d$ orbits happens at R point.
(b) Evolution of energy levels of $\Gamma_5^+$ and $\Gamma_4^-$ at 
R under hydrostatic pressure. Band inversion happens when $a<1.11a_0$.
(c) Electronic band structure with SOC included. A small gap is opened 
in the R-X direction while the Dirac point in the R-M direction is stable 
and protected by crystal symmetry C$_4$ rotation.
}
\end{figure}

Cubic perovskite Cu$_3$PdN is shown in Fig. \ref{fig:crystal_BZ} together
with its BZ and projected surface BZ. The band structure of Cu$_3$PdN is 
shown in Fig.~\ref{fig:bnd}. The valence and conduction bands are 
mainly from Pd-4$d$ (blue) and Pd-5$p$ (red) states, respectively. 
The fatted bands indicate there is band inversion at R with Pd-5$p$ 
lower than Pd-4$d$ by about 1.5 eV. This band inversion is checked 
by hybrid functional~\cite{HSE03, HSE06} calculation. If SOC is 
neglected the occupied and unoccupied low energy 
bands are triply degenerate at R. These states belong to the 
3D irreducible representations $\Gamma_4^-$ and $\Gamma_5^+$ 
of $O_h$ group at R point, respectively. The band inversion 
process in Cu$_3$PdN is shown explicitly in Fig. \ref{fig:bnd}(b), where the
energy levels of $\Gamma_5^+$ and $\Gamma_4^-$ bands at R point 
are calculated under different hydrostatic strains. The band inversion 
happens at $a=1.11a_0$ ($a_0$ is experimental lattice constant) and 
the inversion energy increases as further compressing the lattice.

The intriguing point of the Cu$_3$PdN band structure without SOC 
is that the band crossings due to the band inversion form a nodal ring
because of the coexistence of TRS and IS as addressed in Section 4. 1. 
Due to the cubic symmetry, there are three mutually perpendicular 
nodal rings centering R as schematically shown in Fig.~\ref{fig:crystal_BZ}.

When SOC is considered with spin degree of freedom included, 
the six-fold degenerate states at R are split into one four-fold and 
one two-fold degenerate states. Around the Fermi level, the occupied 
one has $\Gamma_8^+$ symmetry and the unoccupied one has 
$\Gamma_8^-$ symmetry. Both of them are four-fold degenerate.
Along R-X direction, the little group symmetry is lowered to $C_{2v}$. 
The four-fold degenerate states at R are split into two-fold degenerate 
states. The two bands seem to cross each other have the same $\Gamma_5$ 
symmetry and a gap {$\sim$}0.062 eV is opened at the intersection 
as shown in the inset of Fig.~\ref{fig:bnd}(b).

However, along R-M direction, the symmetry is characterized by 
$C_{4v}$ double group. As indicated in {Fig.~\ref{fig:bnd}(c)}, 
the two sets of bands close to Fermi energy belong to 
$\Gamma_7$ and $\Gamma_6$ representation, respectively. 
They are decoupled and the crossing point on R-M path is 
unaffected by SOC.They form a Dirac node near the Fermi energy 
as shown in Fig.~\ref{fig:bnd}(c), which is protected by $C_4$ 
rotational symmetry~\cite{WangZJ_Na3Bi_2012, WangZJ_Cd3As2_2013,XL_Sheng:2014vq}.
If $C_4$ rotational symmetry is broken, the Dirac node 
will be gapped and the Z$_2$ indices of such gapped system are (1;111), 
indicating a strong TI. The band structure of Cu$_3$PdN is 
different from that of antiperovskite Sr$_3$PbO~\cite{FuLiang_antiperovskites_PhysRevB} and Ca$_3$PbO~\cite{Ca3PbO_2012JPSJ},
{where} the band inversion happens at $\Gamma$ point and the involved 
bands belong to the same irreducible presentation, which leads to anti-crossing 
along $\Gamma$-X direction.

\begin{figure}[]
\begin{centering}
\includegraphics[width=0.8\columnwidth]{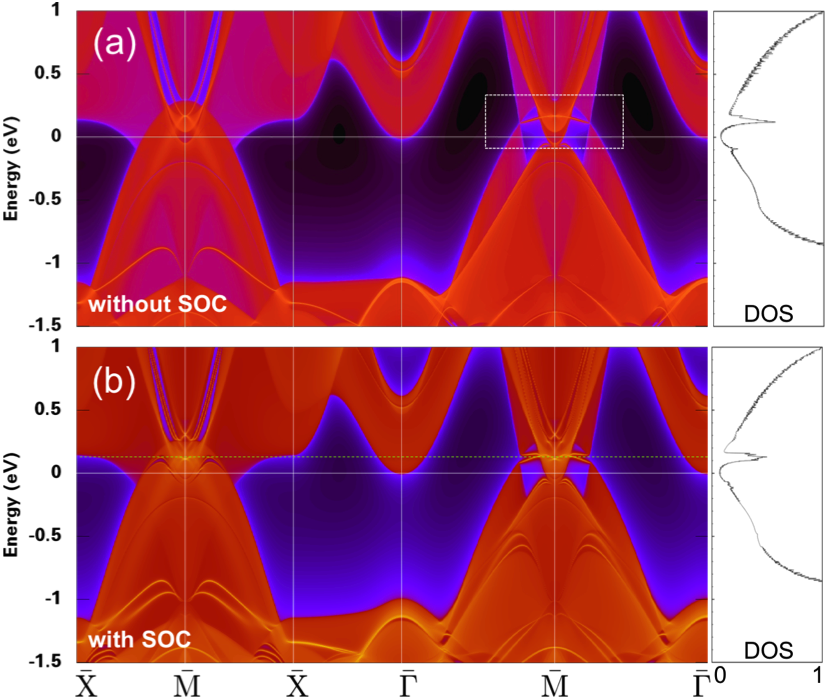}
\par\end{centering}
\protect\caption{\label{fig:sf}(Color online)
Band structures and DOS for (001) surface (a) without and  (b) with SOC .
Without SOC, the nearly flat surface bands are clearly shown in the white dashed box around $\bar{M}$ point.
}
\end{figure}

The band inversion and the 3D Dirac cones in Cu$_3$PdN suggest 
the presence of topologically nontrivial surface states.
The calculated band structures and surface density of states (DOS) 
on semi-infinite (001) surface are presented in Fig.~\ref{fig:sf}.

Without SOC, the bulk state is the same as MTC~\cite{allcarbon_nodeLine2014} 
and there exists surface flat bands nestled inside the projected nodal 
line ring on the (001) surface, namely the ``drumhead" states as shown 
in {Figure \ref{fig:sf}(a)}. The peak-like DOS from these nearly 
flat bands is also clearly shown.
The small dispersion of this ``drumhead" state comes from the fact that 
the nodal line ring is not necessarily on the same energy level due to 
{the particle-hole asymmetry}~\cite{allcarbon_nodeLine2014,Burkov_Topological_nodal_semimetals_2011PRB,Heikkila_2015arXiv}.
{Such 2D flat bands and nearly infinite DOS are proposed as a route to 
achieving high-temperature superconductivity~\cite{PhysRevB.83.220503,2014arXiv1409.3944V,Heikkila_2015arXiv}. }

\begin{figure}[]
\begin{centering}
\includegraphics[width=0.8\columnwidth]{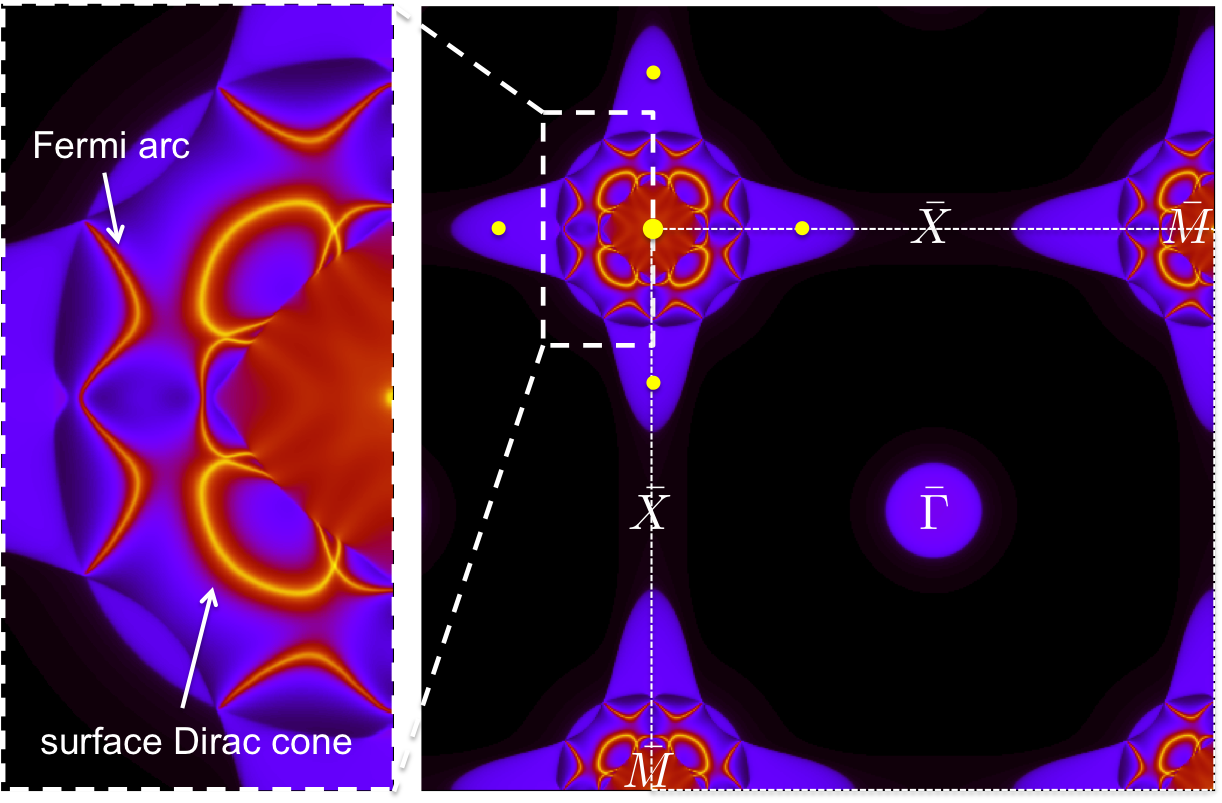}
\par\end{centering}
\protect\caption{\label{fig:SS}(Color online)
{Fermi surface of (001) surface} shown in Fig.~\ref{fig:sf}(b) 
with chemical potential at 0.12 eV (green dashed line in Fig. \ref{fig:sf} (b). 
The yellow dots are projected Dirac points. The bigger one at $\bar{M}$ 
means there are two Dirac points superposed there.
}
\end{figure}

In the presence of SOC, each ring is driven into one pair of Dirac nodes. 
The (001) surface state band structure in Fig.~\ref{fig:sf}(b)
clearly {shows}  the gapped bulk state along $\bar{\Gamma}$-$\bar{M}$ direction 
and the existence of  surface Dirac cone due to {topologically} nontrivial Z$_2$ indices
as seen in Na$_3$Bi~\cite{WangZJ_Na3Bi_2012} and Cd$_3$As$_2$.~\cite{WangZJ_Cd3As2_2013}
The bulk band structure along R-X and R-M overlap each other when projected onto 
(001) surface along the $\bar{X}$-$\bar{M}$ path. The bulk Dirac cones are 
hidden {by} other bulk bands. Therefore, it is difficult to identify the detailed 
connection of Fermi arcs in the Fermi surface plotting as shown in Fig.~\ref{fig:SS}, 
though some eyebrow-like Fermi arcs can be clearly seen around these projected Dirac 
nodes.

{\large\bf 5. Discussion and Prospect}
\hspace*{2pt}


Topological semimetals, as a nontrivial extension of topological classification of 
electronic quantum state from insulator to metal, have been a spotlight in the field
of condensed matter physics in recent years. The theoretical prediction of these
topologically nontrivial quantum state has played a crucial role in stimulating
this field greatly. An uncompleted survey on the existing theoretical proposals
of candidate compounds for each member of TSMs have been listed in Table.~\ref{TSM_proposals}.
The DSM classification is following those of Ref.~\onlinecite{yang_classification_2014}. 
Presently widely studied DSM materials Na$_3$Bi and Cd$_3$As$_2$
are both predicted firstly and then confirmed experimentally. The first WSM of TaAs 
family is also predicted by theoretical calculation. and recently they are under intensive 
study. The proposed FM half-metal states in HgCr$_2$Se$_4$ has also been confirmed 
and it is now much close to the final confirmation of its double WSM state. NLSM 
is the next challenge to be discovered and the predicted candidates are waiting for 
their experimental verifications. 

As mentioned in the introduction, the most interesting phenomena cause by the existence 
of the Weyl points near the Fermi energy is so called chiral anomaly, where the electrons 
under parallel magnetic and electric field will be pumped from one Weyl nodes to another 
one with opposite chirality. In solid states, this pumping process will be eventually balanced 
by the inter-valley scattering terms generated by the impurities leading to a steady state carrying 
extra chiral current along the direction of the magnetic field with the strength being promotional 
to the inner product of E and B fields. How chiral pumping effect can manifests itself in experimental 
measurable quantities is one of the key issues in the study of TSM now. Both theoretical analysis 
and experimental studies seem to suggest that the chiral anomaly will cause universal negative 
magneto resistance being proportional to -B$^2$ only in the weak magnetic field limit. 
In the quantum limit under strong magnetic field, the inter Valley scattering rate will depend 
on field strength as well, which make the behavior of magneto resistance in quantum limit 
to be related to the detail properties of the impurity potential and non universal. Although 
the negative magneto resistance under parallel magnetic and electric fields has been 
experimentally observed in several Weyl, Dirac and even massive Dirac materials, 
how this can be explained uniquely by chiral anomaly is still far from clear. Other 
effects that may induced by chiral anomaly, i.e. chiral magnetic effect, nonlocal 
transport effect and unique optical properties, are still to be observed experimentally.

\begin{figure}[]
\begin{centering}
\includegraphics[width=0.8\columnwidth]{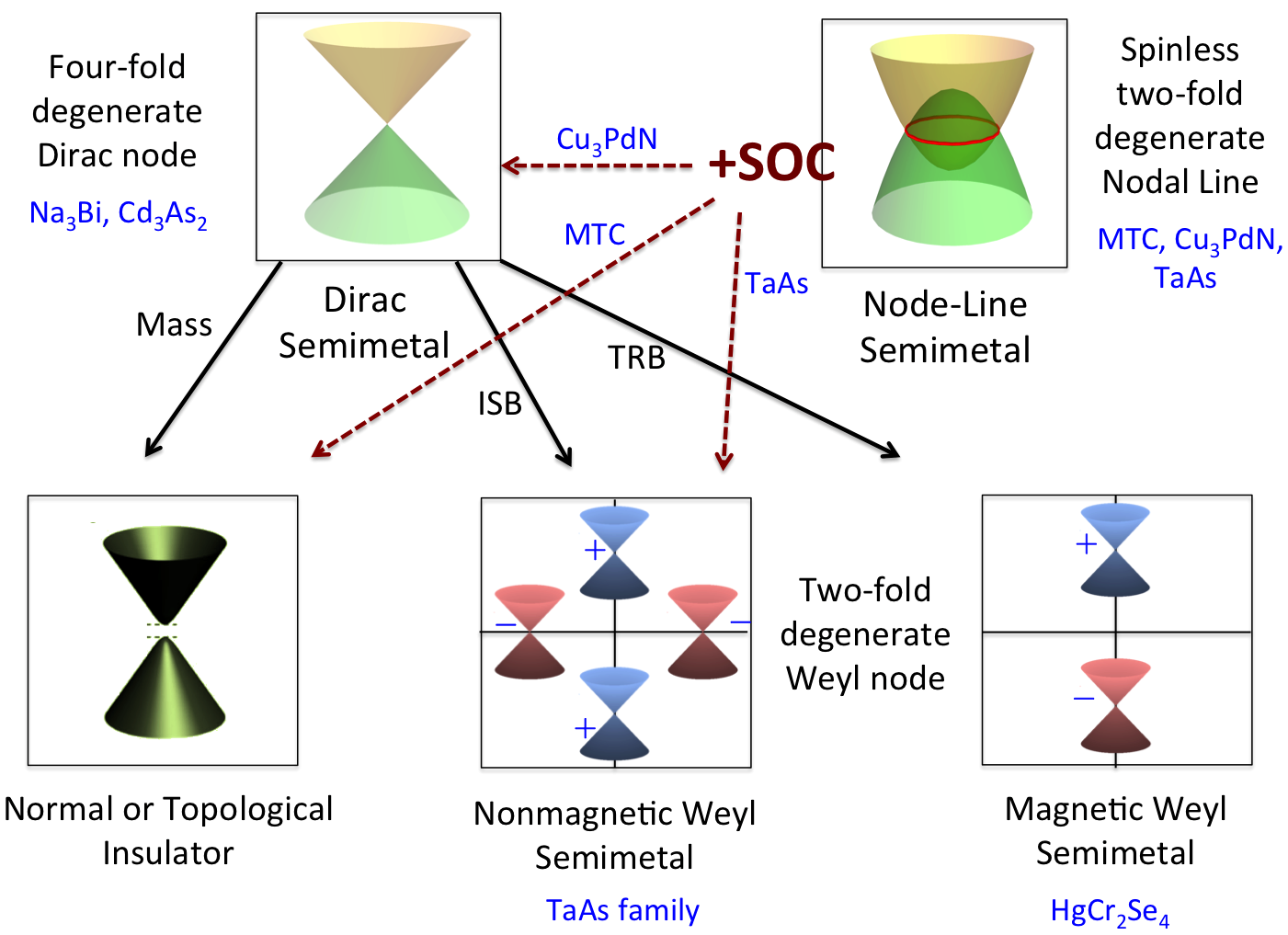}
\par\end{centering}
\protect\caption{
\label{fig:TSM_family}
(Color online) Topological Semimetal (TSM) family and their 
relationship with each other. 
}
\end{figure}

Through this topical review, the relationship among these TSMs is revealed and
summarized in Fig.~\ref{fig:TSM_family}. In the spinless case without including
SOC, the coexistence of TRS and IS protects the NLSM state where the
band inversion leads to nodal points. Starting from such NLSM, including SOC
might lead it to DSM (such as Cu$_3$PdN), TI (such as MTC) and WSM (such as TaAs family).
For DSM, introducing different mass term can drive it to normal insulator or TI and 
breaking either TRS or IS will lead to WSM.

\newcolumntype{C}[1]{>{\centering\arraybackslash}p{#1}}
\begin{table*}%
\caption{An uncompleted survey of existing proposals for compounds hosting TSM.}
\label{TSM_proposals}
\begin{tabular}{|c|c|c|C{3.5cm}|C{3cm}|C{3cm}|}

\hline 
\multicolumn{2}{|c|}{} & Type & Cadidate Compound & Experimental Confirmation & Teoretical Proposal\tabularnewline
\hline 
\multirow{23}{*}{TSM} & \multirow{9}{*}{DSM} & \multirow{5}{*}{Class I} & Na$_{3}$Bi & Ref.~\onlinecite{ChenYL_Na3Bi_2014Science, Xu2013, xu_observation_2014,Ong_Na3Bi_2015, Xiong_Science_2015} & Ref. ~\onlinecite{WangZJ_Na3Bi_2012}\tabularnewline
\cline{4-6} 
 &  &  & Cd$_{3}$As$_{2}$ & Ref.~\onlinecite{ChenYL_Cd3As2_2014NatMa,neupane_observation_2013,PhysRevLett.113.027603,jeon_landau_2014,ZhouXJ_Cd3As2_2014,PhysRevLett.113.246402,Liang2015} & Ref. ~\onlinecite{WangZJ_Cd3As2_2013}\tabularnewline
\cline{4-6} 
 &  &  & BaAuBi-family &  & Ref.~\onlinecite{PhysRevB.91.205128,Wan2015}\tabularnewline
\cline{4-6} 
 &  &  & LiGaGe-family &  & Ref.~\onlinecite{PhysRevB.91.205128}\tabularnewline
\cline{4-6} 
 &  &  & Cu$_{3}$PdN &  & Ref.~\onlinecite{Cu3NPd_Yu}\tabularnewline
\cline{3-6} 
 &  & \multirow{4}{*}{Class II} & BiO$_{2}$ &  & Ref.~\onlinecite{young_dirac_2012}\tabularnewline
\cline{4-6} 
 &  &  & Distorted spinel &  & Ref. ~\onlinecite{PhysRevLett.112.036403}\tabularnewline
\cline{4-6} 
 &  &  & HfI$_{3}$-family &  & Ref.~\onlinecite{PhysRevB.91.205128}\tabularnewline
\cline{4-6} 
 &  &  & TIMo$_{3}$Te$_{3}$-family &  & Ref.~\onlinecite{PhysRevB.91.205128}\tabularnewline
\cline{2-6} 
 & \multirow{5}{*}{WSM} & \multirow{2}{*}{Magnetic} & $R$$_{2}$Ir$_{2}$O$_{7}$ &  & Ref.~\onlinecite{ WanXG_WeylTI_2011}\tabularnewline
\cline{4-6} 
 &  &  & HgCr$_{2}$Se$_{4}$ &  & Ref.~\onlinecite{ XuGang_HgCrSe_2011_PRL}\tabularnewline
\cline{3-6} 
 &  & \multirow{3}{*}{Non-magnetic} & TaAs-family & Ref.~\onlinecite{PhysRevX.5.031013,TaAs_NatPhys2015,PhysRevX.5.031023,PhysRevLett.115.217601,Yang2015, Liu2015aj, Shekhar2015,Jia2015PRB, Xu2015a,Xu2015b,Xu2015c,Xu2015d, 2015arXiv150703983X,FengDL2015} & Ref.~\onlinecite{TaAs_Weng, HuangSM_Weyl}\tabularnewline
\cline{4-6} 
 &  &  & Se or Te under pressure &  & Ref.~\onlinecite{SeTe}\tabularnewline
\cline{4-6} 
 &  &  & $A$Bi$_{1-x}$Sb$_{x}$Te$_{3}$ &  & Ref.~\onlinecite{WS_Vanderbilt_2014} \tabularnewline
\cline{2-6} 
 & \multirow{9}{*}{NLSM} & \multirow{6}{*}{Without-SOC} & Carbon allotropes &  & Ref.~\onlinecite{allcarbon_nodeLine2014,carbon_fanzhang}\tabularnewline
\cline{4-6} 
 &  &  & Cu$_{3}$$R$N &  & Ref.~\onlinecite{ Cu3NPd_Yu, Cu3NPd_Kane}\tabularnewline
\cline{4-6} 
 &  &  & LaN &  & Ref.~\onlinecite{LnX}\tabularnewline
\cline{4-6} 
 &  &  & Black Phosphorus under pressure &  & Ref.~\onlinecite{BP_Zhao2015}\tabularnewline
\cline{4-6} 
 &  &  & Ca$_{3}$P$_{2}$ &  & Ref.~\onlinecite{Ca3P2}\tabularnewline
\cline{4-6} 
 &  &  & CaAg$X$ &  & Ref.~\onlinecite{CaAgX}\tabularnewline
\cline{3-6} 
 &  & \multirow{3}{*}{With-SOC} & PdTaSe$_{2}$\footnotemark[1] &  & Ref.~\onlinecite{PbTaSe2a,PbTaSe2}\tabularnewline
\cline{4-6} 
 &  &  & TITaSe$_{2}$\footnotemark[1] &  & Ref.~\onlinecite{TlTaSe2}\tabularnewline
\cline{4-6} 
 &  &  & SrIrO$_{3}$ &  & Ref.~\onlinecite{CFang_NLSM_PRB}\tabularnewline
\hline 

\end{tabular}
\footnotetext[1]{These candidates have other bulk Fermi surface mixed with proposed Node-Lines.}
\end{table*}

{\large\bf 6. Acknowledgement}
\hspace*{2pt}

H.M.W., X.D. and Z.F. are supported by the National Science Foundation of China,
the 973 program of China (No. 2011CBA00108 and 2013CB921700), and the 
``Strategic Priority Research Program (B)" of the 
Chinese Academy of Sciences (No. XDB07020100).

%
{\large\bf 7. References}
\hspace*{2pt}

\bibliography{section_references}
\end{document}